# *Introducing Dynamic Behavior in Amalgamated Knowledge Bases*

Elisa Bertino

*Dipartimento di Scienze dell'Informazione, University of Milano*
*Via Comelico 39/41, 20135 Milano, Italy*
*e-mail: bertino@dsi.unimi.it*

Barbara Catania

*Dipartimento di Informatica e Scienze dell'Informazione, University of Genova*
*Via Dodecaneso 35, 16146 Genova, Italy*
*e-mail: catania@disi.unige.it*

Paolo Perlasca

*Dipartimento di Scienze dell'Informazione, University of Milano*
*Via Comelico 39/41, 20135 Milano, Italy*
*e-mail: perlasca@dsi.unimi.it*

**Abstract**

The problem of integrating knowledge from multiple and heterogeneous sources is a fundamental issue in current information systems. In order to cope with this problem, the concept of *mediator* has been introduced as a software component providing intermediate services, linking data resources and application programs, and making transparent the heterogeneity of the underlying systems. In designing a mediator architecture, we believe that an important aspect is the definition of a formal framework by which one is able to model integration according to a declarative style. To this purpose, the use of a logical approach seems very promising. Another important aspect is the ability to model both static integration aspects, concerning query execution, and dynamic ones, concerning data updates and their propagation among the various data sources. Unfortunately, as far as we know, no formal proposals for logically modeling mediator architectures both from a static and dynamic point of view have already been developed. In this paper, we extend the framework for amalgamated knowledge bases, presented in (Subrahmanian, 1994), to deal with dynamic aspects. The language we propose is based on the Active U-Datalog language (Bertino *et al.*, 1998), and extends it with annotated logic and amalgamation concepts from (Kifer and Subrahmanian, 1992; Subrahmanian, 1987). We model the sources of information and the mediator (also called *supervisor*) as Active U-Datalog deductive databases, thus modeling queries, transactions, and active rules, interpreted according to the PARK semantics (Gottlob *et al.*, 1996). By using active rules, the system can efficiently perform update propagation among different databases. The result is a logical environment, integrating active and deductive rules, to perform queries and update propagation in an heterogeneous mediated framework.

## 1 Introduction

The problem of integrating knowledge from multiple and heterogeneous sources is a crucial issue in current information system technology. Very often the knowledge required to perform a certain task is factored in several heterogeneous systems and specific tools are needed in order to acquire, store, manage, query and update data and knowledge in an integrated way.

The development of an integrated information system entails addressing different problems, ranging from the differences in hardware/software platforms, to heterogeneity in the database management systems (DBMS), to semantic data heterogeneity, to operational issues such as update propagation and consistency maintenance for related information. Solutions to these problems are provided by efforts in different areas (Bukhres and Elmagarmid, 1996; Schek *et al.*, 1993; Elmagarmid, 1993; Ceri and Widom, 1993; Do and Drew, 1995; Gupta *et al.*, 1993; Chawathe *et al.*, 1996; Subrahmanian, 1994).

In order to cope with the management of heterogeneous systems, the concept of *mediator* has been introduced (Wiederhold, 1992). A mediator can be defined as a software system component providing intermediate services, linking data resources and application programs. In the context of heterogeneous knowledge bases, mediators provide users with an integrated view of multiple sources, making transparent the underlying data heterogeneity. The central problem in mediation is the identification of relevant resources for the client model and the retrieval of relevant data at the time of a client inquiry. This goal is achieved by three main functions: selection of data, translation of the user query into queries suitable for the underlying sources, merging of the resulting data by removing redundancy and inconsistency. Through this entire process, the user should not be aware of the underlying heterogeneity.

In designing a mediator architecture, we believe that an important aspect is the definition of a formal framework by which it is possible to model integration among heterogeneous systems according to a declarative style. Such a framework may be quite useful in understanding how and when information has to be integrated together. In general, this is a very difficult task, especially when the number of involved systems is large. This consideration calls for a declarative approach allowing one to fully model all aspects of integration and to provide the basis by which properties of heterogeneous information systems can be proved. To this purpose, the use of a logical approach seems very promising.

Among the logical approaches that have been proposed (Bowen and kowalski, 1982; Fagin *et al.*, 1983; Fagin *et al.*, 1986; Grant *et al.*, 1991; Subrahmanian, 1989; Subrahmanian, 1994; Whang *et al.*, 1991; Zicari *et al.*, 1991), the *amalgamated knowledge base* framework proposed by Subrahamanian (Subrahmanian, 1994) is one of the few proposals providing a formal logical foundation to cooperative knowledge bases. It also represents the formal basis of the HERMES system (Subrahmanian *et al.*, 1996). In this framework, generalized annotated logic (Kifer and Subrahmanian, 1992; Subrahmanian, 1987) is used to model data sources and the notion of supervisor is introduced as a *mediator*, amalgamating the knowledge coming from the local databases. The use of annotated logic provides the right formalism for modeling knowledge at different degrees of inconsistencies and uncertainties, typical of an heterogeneous environment. A model theoretic and a fixpoint semantics for the proposed framework have also been proposed, thus leading to the definition of a fully declarative approach for modeling amalgamated knowledge bases.

Even if the amalgamated knowledge based framework is quite appealing, it lacks the modeling of dynamic aspects. In particular, both the local databases and the supervisor have no dynamic behavior. This is a strong limitation since it means that the system is not able to react to events. Indeed, we believe that, in order to provide most of the functionalities required to support heterogeneous knowledge bases, mediators should implement two different types of integration:

- *Static integration*. With static integration we mean the ability to model heterogeneous sources of data, intended as different databases storing (possibly related) data and intensional knowledge on data, and the ability to integrate several data management systems to collectively provide information to answer user queries.
- *Dynamic integration*. With dynamic integration we mean the ability to update data and to propagate updates among the various data sources.
  Here, the main issue concerns how knowledge can be modified in an integrated way, introducing new information inside the local databases through the use of the mediator, and, at the same time, how consistency of the knowledge bases can be guaranteed. These aspects can be supported through the use of *active rules*.

In this paper we extend the amalgamated knowledge base framework presented in (Subrahmanian, 1994) to deal with logical languages modeling updates and active rules. The language we propose is based on the Active U-Datalog language (Bertino *et al.*, 1998) and extends it with annotated logic and amalgamation concepts taken from (Kifer and Subrahmanian, 1992; Subrahmanian, 1987). The result is a logical framework modeling both static and dynamic aspects in integrating knowledge from multiple heterogeneous sources. Dynamic integration is then provided through updates and active rules, both at the local and global level.

The Active U-Datalog language is a language for integrating active rules, deductive rules and updates in a uniform logical context; it is based on Update Datalog (U-Datalog for short) (Bertino *et al.*, 1997) and extends it with support for active rules, in the style of the PARK semantics (Gottlob *et al.*, 1996). In Active U-Datalog, update atoms appear in rule bodies; the execution of a goal (also called a *transaction*) is based on a deferred semantics, by which several updates are generated from predicate evaluation, but not immediately executed; rather, they are collected and are executed only at the end of the query-answering process. In Active U-Datalog, updates are expressed by using constraints. For example, $+p(a)$ states that in the new state $p(a)$ must be true whereas $-p(a)$ states that in the new state $p(a)$ must be false. Each atomic solution generates a set of updates.

Active rules allow several dynamic aspects to be represented, such as transaction execution, reactive behavior and, more specifically, update propagation, in a uniform logical framework. The semantics proposed for active rules is based on the PARK semantics (Gottlob *et al.*, 1996). The PARK semantics has been designed with the intent of overcoming the limitations of previously defined semantics for active rules. In particular, given a set of ECA (Event-Condition-Action) rules, that is rules of the form "ON event IF condition THEN action", the PARK semantics satisfies several properties. First of all, it is non-ambiguous, that is, it always guarantees execution confluence. Moreover, it is flexible with respect to conflict resolution.

A conflict is a situation where two or more active rules can be fired and one of these rules requires the insertion of an atom $a$ in the database, whereas at least one of the others requires the deletion of $a$ from the database. A conflict resolution policy is a method to determine which actions should be executed in presence of a conflict and which others should be suppressed. Under the PARK semantics, the conflict resolution policy can be chosen according to specific application requirements. For example, the policy can specify that insertions must always prevail upon deletions. A fixpoint semantics is used to determine the result of the application of a set of active rules. The proposed semantics guarantees the termination of the evaluation process. The use of the PARK semantics allows the system to handle updates generated by deductive rules and updates generated by active rules in a uniform way. It is important to note that, in the context of the paper, the term consistency refers to a situation in which conflicts are avoided and it is therefore different from the general concept of consistency used in the context of integrity constraint checking.

The logical framework we introduce in this paper relies on the Active U-Datalog logical language to model both the local knowledge bases and the mediator. More precisely, we introduce the concept of *Amalgamated Active-U-Datalog knowledge base* as a knowledge base supporting the following features:

1. *Multiple sources of data (local databases)*. We model each deductive database as an Active U-Datalog database, extended with annotated logic concepts. Thus, each local database consists of an extensional database (i.e., a set of facts), an intensional database represented by a set of deductive Active U-Datalog rules, and a set of active Active U-Datalog rules. Atoms in deductive and active rules are annotated with values taken from a given complete lattice of truth values. The use of Active U-Datalog provides the ability to model, not only queries, but also updates inside each database. Moreover, active rules allow the local databases to react to external events (in our context, represented by updates). The resulting language can be thought as an interface language by which local source knowledge is represented, for example through a wrapper-based approach, before integration.
2. *Integration of local databases*. Local databases can be integrated through a *supervisor*. As defined in (Subrahmanian, 1994), a supervisor specifies a set of rules by which the local knowledge can be integrated. The local databases, the supervisor and a set of axioms required to gather information from different databases form the *amalgam*. Differently from what has been presented in (Subrahmanian, 1994), in our framework the supervisor, which is an Amalgamated Active U-Datalog program, can execute updates against local databases and support active rules. Such rules can propagate updates depending on the whole status of the integrated system.

For both aspects, a fixpoint semantics is proposed, integrating those presented in (Subrahmanian, 1994) and in (Bertino *et al.*, 1998).

We recall that other approaches have been proposed for update propagation in the context of heterogeneous databases (Ceri and Widom, 1993; Do and Drew, 1995; Gupta *et al.*, 1993). However, most of the proposed approaches suggest how to use active rules to perform specific tasks, such as schema integration and integrity constraint checking. On the contrary, in the proposed framework, active rules are introduced at a general level. Moreover, the use of the PARK semantics provides a clear integration of active and deductive rules and makes the approach much more flexible with respect to the problem of conflict resolution. Indeed, differently from other proposals, where conflicts are solved by assigning priority to rules (Hanson, 1996; Stonebraker *et al.*, 1990; Widom and Finkelstein, 1990), the PARK semantics allows the application programmer to choose the best conflict resolution policy to apply in a particular case. In solving a conflict, information about the structure and the current state of the system can be considered. This is particularly important in an integration framework, where update propagation often depends on information distributed among various databases.

The paper is organized as follows. In Subsection 1.1 we compare our approach with several existing approaches for mediating heterogeneous sources. The syntax of the proposed framework is presented in Section 2. In Sections 3, 4, and 5 we then introduce the semantics of Amalgamated Active U-Datalog, together with several examples of its application. Finally, Section 6 presents some conclusions and outlines future work.

### *1.1 Related Work*

When dealing with the integration of heterogeneous information sources, two important issues concern *semantic interoperability*, intended as the capability of representing local data, defined according to a local data model, in terms of a common data model, and of providing an integrated view of local schemas, and *data modification capabilities*, intended as the ability to update data and to propagate updates among the various data sources.

In the literature, several approaches for mediating heterogeneous sources have been proposed (Arens *et al.*, 1996a; Arens *et al.*, 1993; Arens *et al.*, 1996b; Garcia-Molina *et al.*, 1997; Beeri *et al.*, 1997; Levy *et al.*, 1996; Subrahmanian *et al.*, 1996; Eiter *et al.*, 1999) but most of them only deal with the semantic interoperability problem and do not address data modification capabilities. Among those approaches, we recall the *SIMS* approach (Arens *et al.*, 1996a; Arens *et al.*, 1993; Arens *et al.*, 1996b), which is based on the creation of a *global domain model* used to represent the various local information sources. Queries are expressed in such global domain model and, in order to process them, an optimized query plan is built by rewriting global domain queries in terms of local sources queries, performing a semantic query optimization. The *TSIMMIS* approach (Garcia-Molina *et al.*, 1997) addresses the heterogeneous source integration problem by considering a mediator network composed of mediators and wrappers. In this network, each mediator is able to use local information sources through wrappers and/or through other mediators. Wrappers are responsible for converting global queries into local source queries. The *Information Manifold* system (Levy *et al.*, 1996) has been developed with the aim of retrieving heterogeneous information over the Web. It provides a uniform access to information sources by using a declarative description of both the content and the query capabilities of such sources. The descriptions of information sources are stored in the so called "capability records", that are used to build efficient local query plans. The meaning of capability records is similar to that of "yellow page servers" proposed in the context of the *HERMES* system (Subrahmanian *et al.*, 1996). The aim of such a system is that of providing an environment for defining mediators. It is based on the *Hybrid Knowledge Base* theory (Lu *et al.*, 1996) for integrating information belonging to different data sources. It provides a general declarative language and a specific set of tools whose purpose is to make easier the steps involved in the creation of mediators. Global queries submitted to one of such mediators may also trigger actions based on the analysis of the answer produced by such queries.

Related to the general problem of rewriting queries, there is the problem of rewriting queries using views, that is particularly relevant in data mining and data warehouse contexts. Such problem can be stated as follows: given a query and a set of views, find a new query, equivalent to the given one, that uses only the given set of views. In (Beeri *et al.*, 1997) this problem has been solved by modeling rewritings in description logics (Borgida, 1995), showing that, under particular conditions, the considered problem is decidable in polynomial time.

Unlike the above approaches, with the exception of the HERMES one, our approach focuses on dynamic aspects. We assume that local data sources are expressed as Active U-Datalog databases. Thus, Active U-Datalog can be seen as the interface language between local sources and the mediator. Local Active U-Datalog databases are then integrated by a special mediator, called *supervisor*, able to retrieve and manage local information. The main feature of our system is that both local databases and supervisor are characterized by an active behavior intended as the ability of automatically reacting to external events (in our context, represented by updates) arising in the system.

Among the above approaches, only HERMES supports dynamic aspects. However, in our approach, actions can be triggered both at the local and at the global levels, respectively by local databases and the supervisor. By contrast, in HERMES actions can only be activated at global level and no support for local activation is provided.

Finally, some relationships exist also between our work and agent technology. An agent can be defined as "a self-contained program capable of controlling its own decision-making and acting, based on its perception of its environment, in pursuit of one or more objectives" (Jennings and Woldridge, 1996). An agent should be characterized by several properties such as, for example, the ability to interact with other agents and to react to changes of its state or arising in the overall environment. The model we have developed partly satisfies the above properties and has therefore some similarities with agent technology. In our model, each local database is characterized by a state and by a set of deductive and active rules. As such, our local databases are self-contained and able to modify their own state. Moreover, communication among local databases is indirectly supported via the supervisor since it has a global visibility of the whole system and can use this knowledge to answer queries and to perform modifications involving the whole system. However, differently from agent systems, the proposed framework has been cast in a specific environment, with the aim of providing a formal approach for analyzing heterogeneous environments, both from a static and dynamic point of view. This is different from agent technology, since in that case methods are provided to deal with arbitrary agents, used to perform specific tasks in arbitrary environments.

## 2 Amalgamated Active-U-Datalog: the syntax

In order to support a logical framework modeling cooperation and integration of knowledge from different and heterogeneous databases, Subrahamanian proposed a framework, based on annotated logic, for amalgamating the knowledge contained in several heterogeneous local databases (Subrahmanian, 1994). Each local database is a generalized annotated program (GAP) (Kifer and Subrahmanian, 1992; Subrahmanian, 1987), that is, a logic program whose semantics is interpreted over a complete lattice[1] of truth values. Some examples of lattices are (Subrahmanian, 1994):

- the lattice TWO, containing the classical truth value $true$ and $false$, with $false$ lower than $true$;
- the lattice FOUR; in such lattice, $\top$ represents the truth value *inconsistent*, $\bot$ represents the truth value *unknown*, and $t$ and $f$ represent the usual values $true$ and $false$, respectively; by denoting with $<$ the ordering existing between lattice values, the following relationships hold: $\bot < true$, $\bot < false$, $false < \top$, $true < \top$;
- the lattice $\mathcal{R}[0, 1]$, set of real numbers between 0 and 1;
- the lattice $TIME_1$, representing the power-set of non-negative integers, ordered under $\subseteq$;
- the lattice $TIME_2$, representing the set of all closed intervals of non-negative real numbers, ordered under $\subseteq$.

As we have already remarked, the proposed framework does not model dynamic aspects, such as updates and active rules. In order to overcome this limitation, in the following, we extend the amalgamation theory to cope with updates and active rules. In order to do that, we apply amalgamation theory to local sources represented as Active U-Datalog databases (Bertino *et al.*, 1998). Since Active U-Datalog programs support update representation and execution, the amalgamation of Active U-Datalog programs allows us to combine not only the static knowledge of each single database but also the dynamic one, represented by updates. The resulting language is called *Amalgamated Active U-Datalog*.

In the following, we first describe how lattice values can be inserted inside Active U-Datalog programs, modeling *local databases*. The resulting language is called *Annotated Active U-Datalog*. Then, we show how the static and dynamic information contained in the local databases can be amalgamated, resulting in the *Amalgamated Active U-Datalog* framework. In defining such new framework, we consider as source for the truth values a complete lattice $\mathcal{T}$.

### 2.1 Annotated Active U-Datalog

An Annotated Active U-Datalog program is a logical program modeling both deductive rules and active rules. Both deductive and active rules are defined by using deductive and update atoms. Such atoms can be annotated with the truth values chosen from a given complete lattice $\mathcal{T}$.

In order to introduce the Annotated Active U-Datalog language, we first define the concept of annotation, then we introduce the concept of annotated atom, and finally we describe the rules that can be represented in Annotated Active U-Datalog.

---

[1] A lattice is a partially ordered set where all finite subsets have a least upper bound and a greatest lower bound. In a complete lattice, the least upper bound and a greatest lower bound also exist for infinite subsets. Every finite lattice is complete.

*2.1.1 Annotations*

Given a complete lattice $\mathcal{T}$, annotations are constructed upon a specific language, called $\mathcal{T}$-language. Based on this language, an annotation can be either a syntactic representation of the lattice elements or it is obtained by applying a computable function to such elements. The representation domain of the $\mathcal{T}$ language is therefore the structure corresponding to the $\mathcal{T}$ lattice.

*Definition 1 ($\mathcal{T}$-language)*
Let $\mathcal{T}$ be a complete lattice of truth values. The $\mathcal{T}$-language $(\mathcal{C}, \mathcal{F}, \mathcal{V})$, used to represent annotations over $\mathcal{T}$, is composed of:

- a set $\mathcal{C}$ of constant symbols;
- an union $\mathcal{F}$ of sets $\mathcal{F}^i$ of total continuous[2] functions ($\bigcup_{i \geq 1} \mathcal{F}^i$), each of type $(\mathcal{T})^i \to \mathcal{T}$, called *annotation functions over lattice $\mathcal{T}$*, such that:
  — each $f \in \mathcal{F}^i$ is computable;
  — each set $\mathcal{F}^i$ contains an $i$-ary function $\sqcup_i$[3] that, given $\mu_1, \ldots, \mu_i$ as input, returns the *least upper bound* (*lub*) of $\{\mu_1, \ldots, \mu_i\}$.

Moreover, we assume that the $\mathcal{T}$-language contains an infinite set $\mathcal{V}$ of variable symbols. □

We now introduce the concept of *annotation*.

*Definition 2 (Annotation)*
Let $\mathcal{T}$ be a complete lattice of truth values. Let $T = (\mathcal{C}, \mathcal{F}, \mathcal{V})$ be a $\mathcal{T}$-language. An *annotation* $\psi$ over $T$ is a term constructed over $T$ (T-term). A T-term $t$ can be either:

- a simple annotation term, if $t \in \mathcal{C}$ or $t \in \mathcal{V}$;
- a complex annotation term, if $t = f(\mu_1 \ldots, \mu_i)$, where $f \in \mathcal{F}^i$ and $\mu_1 \ldots, \mu_i$ are annotation terms. □

In the following, when no otherwise specified, we consider as $\mathcal{T}$-language the language in which the constants coincide with the elements of $\mathcal{T}$.

*Example 1*
Let $\mathcal{T}$ be the lattice $\mathcal{R}[0, 1]$, $T = (\mathcal{C}, \mathcal{F}, \mathcal{V})$ a $\mathcal{T}$-language such that:

- $\mathcal{C}$ is a syntactic representation of the considered real numbers;
- $\mathcal{F} \equiv \mathcal{F}^2$ is composed of a binary function $f : (\mathcal{T})^2 \to \mathcal{T}$ returning the minimum value between its arguments and a binary function $\sqcup_2 : (\mathcal{T})^2 \to \mathcal{T}$ returning the least upper bound between its arguments;
- $\mathcal{V} \equiv \{X_1, \ldots, X_n, \ldots\}$.

Then, $0.75$ and $X_1$ are simple annotation terms whereas $f(0.25, 0.75)$ and $\sqcup_2(0.5, 0.75)$ are complex annotation terms. ◇

---

[2] Hence monotonic.
[3] For simplicity, when there is not ambiguity, we use $\sqcup$ instead of $\sqcup_i$.

*2.1.2 Atoms*

The following definition introduces the language over which atoms are constructed.

*Definition 3* (*Base language*)
Let $T = (\mathcal{C}, \mathcal{F}, \mathcal{V})$ be a $\mathcal{T}$-language. Let $\Sigma = \{\Sigma_c, \Sigma_a\}$ be a many sorted signature, such that $\Sigma_c$ is a set of constant value symbols, and $\Sigma_a$ is $\mathcal{F} \cup \mathcal{C}$. Sets $\Sigma_c$ and $\Sigma_a$ are disjoint. Let $\Pi$ be a set of predicate symbols, partitioned into extensional predicate symbols $\Pi^e$, intensional predicate symbols $\Pi^i$, and update predicate symbols $\Pi^u$. We assume that $\Pi^u = \{+p, -p \mid p \in \Pi^e\}$. Let $V$ be a family of sets of variable symbols for each sort $V = \{V_c, V_a\}$, $V_a = \mathcal{V}$. $(\Pi, \Sigma, V)$ is called *base language*. We denote with $Term_t$ the set $\Sigma_t \cup V_t$, with $t \in \{c, a\}$. [4]

The base language can be used to construct atoms as follows.

*Definition 4* (*Atoms, annotated atoms*)
Let $T = (\mathcal{C}, \mathcal{F}, \mathcal{V})$ be a $\mathcal{T}$-language and $(\Pi, \Sigma, V)$ a base language. We denote with $(\Pi, \Sigma, V)$-atom an atom whose predicate belongs to $\Pi$ and whose terms are in $\Sigma_c \cup V_c$. More precisely:

- $(\Pi^i, \Sigma, V)$-atoms are called intensional deductive atoms.
- $(\Pi^e, \Sigma, V)$-atoms are called extensional deductive atoms.
- $(\Pi^u, \Sigma, V)$-atoms are called update atoms. Insertions are denoted by update atoms prefixed by $+$, whereas deletions are prefixed by $-$.
- Atoms of the form $A : \psi$, where $A$ is an (intensional, extensional, update) atom and $\psi$ is an annotation over $T$ are called *annotated atoms*. $A : \psi$ is *c-annotated* if $\psi \in \mathcal{C}$, *v-annotated* if $\psi \in \mathcal{V}$, *t-annotated* if $\psi$ is a complex annotation term. The meaning assigned to an annotated update atom $\alpha_i A_i : \mu_i$ is that of inserting/deleting the annotated atom $A_i : \mu_i$.

An annotated literal is an annotated intensional or extensional atom or its negation. An atom (literal) is ground if it does not contain variables. □

*Example 2*
Let $T = (\mathcal{C}, \mathcal{F}, \mathcal{V})$ be the $\mathcal{T}$-language of Example 1 and $(\Pi, \Sigma, V)$ the *base language* defined as follows:

- $\Pi = \{\Pi^e, \Pi^i, \Pi^u\} \equiv \{p_1, \ldots, p_n\} \cup \{q_1, \ldots, q_m\} \cup \{+p_1, -p_1, \ldots, +p_n, -p_n\}$ where the arity of $p_i, 1 \leq i \leq n$, and $q_j, 1 \leq j \leq m$, is 2;
- $\Sigma = \{\Sigma_c, \Sigma_a\} = \{\{a, b, c\}, \mathcal{C} \cup \mathcal{F}\}$;
- $V = \{V_c, V_a\} \equiv \{Y_1, \ldots, Y_q, \ldots\} \cup \{X_1, \ldots, X_s, \ldots\}$.

Then, $p_1(a, b) : 0.5$ is an extensional c-annotated atom, $+p_1(a, b) : X_1$ is an update v-annotated atom, whereas $q_1(b, c) : f(0.25, 0.75)$ and $q_2(Y_1, Y_2) : f(X_1, X_2)$ are intensional t-annotated atoms, the first is ground, the second is not. ◇

In the following, when we do not specify otherwise, we use the term "annotated atom" to refer either a c-annotated, a v-annotated, or a t-annotated atom.

---

[4] In the following we assume that a substitution is a pair of functions $\theta = \{\theta_c, \theta_a\}$, dealing respectively with variables in $V_c$ and $V_a$.

*2.1.3 Rules*

Deductive and active rules are defined as follows.

*Definition 5* (*AAU-Datalog Deductive Rule*)
An AAU-Datalog Deductive Rule is a rule of the form

$$A_0 : \mu_0 \leftarrow A_1 : \mu_1, \ldots, A_k : \mu_k | \alpha_{k+1} A_{k+1} : \mu_{k+1}, \ldots, \alpha_n A_n : \mu_n$$

with $\alpha_j \in \{+, -\}$, $j = k+1, \ldots, n$, and $A_i : \mu_i$, $i = 0, \ldots, n$, annotated atoms such that:

- $A_0 : \mu_0$ is an annotated $(\Pi^i, \Sigma, V)$-atom;
- $A_l : \mu_l$, $l = 1, \ldots, k$, is a c-annotated or v-annotated $(\Pi^i \cup \Pi^e, \Sigma, V)$-atom;
- $A_j : \mu_j$, $j = k+1, \ldots, n$, is a c-annotated or v-annotated $(\Pi^e, \Sigma, V)$-atom;
- $\alpha_s A_s \neq \alpha_t A_t$ for $k+1 \leq s, t \leq n$, $s \neq t$.

$A_0 : \mu_0$ is the head of the rule, $A_1 : \mu_1, \ldots, A_k : \mu_k | \alpha_{k+1} A_{k+1} : \mu_{k+1}, \ldots, \alpha_n A_n : \mu_n$ is the body, $A_1 : \mu_1, \ldots, A_k : \mu_k$ is the query part whereas $\alpha_{k+1} A_{k+1} : \mu_{k+1}, \ldots, \alpha_n A_n : \mu_n$ is the update part; the update and query part cannot be both empty. We require for deductive rules the following safety condition: each variable appearing in the head (or in its annotation) must also appear in a deductive atom in the body of the same rule (or in the annotation of its body atoms). □

The intuitive meaning of a deductive rule

$$A_0 : \mu_0 \leftarrow A_1 : \mu_1, \ldots, A_k : \mu_k | \alpha_{k+1} A_{k+1} : \mu_{k+1}, \ldots, \alpha_n A_n : \mu_n$$

is: "if $A_i$, $i = 1, \ldots, k$, is true with truth value $\mu_i$, then $A_0$ is true with truth value $\mu_0$ and, as side effect, the updates $\alpha_{k+1} A_{k+1} : \mu_{k+1}, \ldots, \alpha_n A_n : \mu_n$ are requested".

*Definition 6* (*AAU-Datalog Active Rule*)
An AAU-Datalog Active Rule is a rule of the form:

$$\alpha_1 A_1 : \mu_1, \ldots, \alpha_k A_k : \mu_k | L_{k+1} : \mu_{k+1}, \ldots, L_n : \mu_n \rightarrow \alpha_{n+1} A_{n+1} : \mu_{n+1}, \ldots,$$
$$\alpha_{n+m} A_{n+m} : \mu_{n+m}$$

with $\alpha_j \in \{+, -\}$, $j = 1, \ldots, k, n+1 \ldots, n+m$, $A_l : \mu_l$, $l = 1, \ldots, k, n+1, \ldots, n+m$, annotated atoms, and $L_h : \mu_h$, $h = k+1, \ldots, n$ annotated literals such that

- $A_i : \mu_i$, $i = 1, \ldots, k$, is a c-annotated or v-annotated $(\Pi^e, \Sigma, V)$-atom;
- $A_i : \mu_i$, $i = n+1, \ldots, n+m$, is an annotated $(\Pi^e, \Sigma, V)$-atom;
- $L_j : \mu_j$, $j = k+1, \ldots, n$, is a c-annotated or v-annotated $(\Pi^i \cup \Pi^e, \Sigma, V)$-literal;
- $\alpha_p A_p \neq \alpha_q A_q$ for $1 \leq p, q \leq k$, $p \neq q$, and $\alpha_s A_s \neq \alpha_t A_t$ for $n+1 \leq s, t \leq n+m$, $s \neq t$.

$\alpha_1 A_1 : \mu_1, \ldots, \alpha_k A_k : \mu_k$ is the event part, $L_{k+1} : \mu_{k+1}, \ldots, L_n : \mu_n$ is the condition part, and $\alpha_{n+1} A_{n+1} : \mu_{n+1}, \ldots, \alpha_{n+m} A_{n+m} : \mu_{n+m}$ is the action part, that cannot be empty. We require for active rules the following safety conditions: each variable occurring in a rule head (or in its annotation) should also occur in the body of the same rule (or in the annotation of some of its atoms); each variable occurring in a negated literal (or in its annotation) in the rule body must also occur in some positive literal (or in its annotation) in the rule body. □

While intensional rules provide deductive power to our framework, active rules allow the system to autonomously react to the current (possibly inconsistent) state and to take appropriate actions in order to assure desired properties on the final state. The intuitive meaning of the active rule

$$\alpha_1 A_1 : \mu_1, \ldots, \alpha_k A_k : \mu_k |\ L_{k+1} : \mu_{k+1}, \ldots, L_n : \mu_n \to \alpha_{n+1} A_{n+1} : \mu_{n+1}, \ldots,$$
$$\alpha_{n+m} A_{n+m} : \mu_{n+m}$$

is: "if the events $\alpha_1 A_1 : \mu_1, \ldots, \alpha_k A_k : \mu_k$ occur and $L_i, i = k+1, \ldots, n$, is true with truth value $\mu_i$, then execute actions $\alpha_{n+1} A_{n+1} : \mu_{n+1}, \ldots, \alpha_{n+m} A_{n+m} : \mu_{n+m}$".

It is important to note that the previous definition prevents a rule from containing the same update atom with two distinct annotations (fourth item of Definitions 5 and 6).

In the following, we call *AAU-Datalog* rule both a deductive and active AAU-Datalog rule.

An *Annotated Active U-Datalog program (or AAU-Datalog)* is finally defined as follows.

*Definition 7* (*AAU-Datalog program*)
An *Annotated Active U-Datalog (or AAU-Datalog) program or database* is composed of:

- a set of c-annotated extensional ground atoms (extensional database *EDB*);
- a set of AAU-Datalog deductive rules (intensional database *EDB*);
- a set of AAU-Datalog active rules (*AR*).

Given an AAU-Datalog program $P$, we denote with $\mathbf{GI(P)}$ all the ground instances of rules in $P$, containing only c-annotations. □

Due to the dynamic properties of an AAU-Datalog program, a deductive rule with no head and with a non-empty query part is called *simple transaction*. Simple transactions are usually preceded by "?", as usual in deductive databases. A *complex transaction* is a sequence of simple transactions $T_1; \ldots ; T_k$, each executed in the state obtained by the execution of the previous transactions (see Section 5).

*Example 3*
Suppose that three sensors monitor the air quality in three distinct places of a given town, by considering the level of a given set of substances. Depending on whether the level of a given substance has a presence above the danger threshold and on how may sensors detect this situation, we want to partially or totally block car circulation. If a sensor detects that the danger threshold for a given substance has been exceeded twice, a critical situation is detected.

Information concerning the three local sensors can be modeled by three AAU-Datalog databases. Partial and total block policies will then be managed by a supervisor (see Example 6). In order to model local databases, we make the following assumption: information concerning the level of a given substance is traced by predicate `sensor`. In particular, the annotated atom `sensor(X): t` means that the level of substance X has a presence over the danger threshold, `sensor(X): f` means that the level of substance X has a presence under the danger threshold, `sensor(X): ⊥` means that we do not know the level of substance X. We assume that such predicate is externally updated. We also assume that partial and total blocks information are represented by the extensional predicates `total_block` and `partial_block`, whose arguments are the substances causing the block.

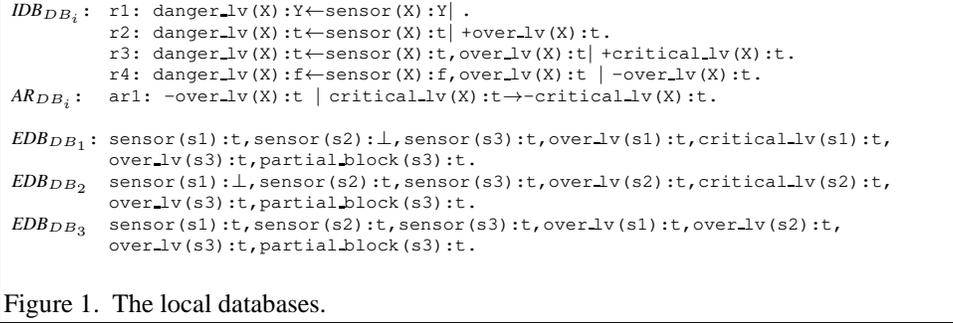

Figure 1. The local databases.

Figure 1 illustrates the local databases. Predicate `danger_lv` specifies whether the danger threshold for a given substance has been exceeded and, at the same time, it updates the extensional database. In particular, the first time a specific substance exceeds the danger threshold, an over level warning is generated, by updating the extensional database. If this happens twice, a critical situation is gained and a critical level warning is generated, updating again the database. The extensional database is also modified when the concentration of the considered substance decreases. In this case, all the warnings are removed. The active rule removes critical warnings when the over level warning is removed.

The local extensional database of $DB_1$ specifies that substance `s1` has exceeded twice its danger threshold and a substance `s3` has exceeded once its danger threshold. A partial block due to substance `s3` has also been detected. No information concerning substance `s2` is known. A similar situation arises in the local database $DB_2$. On the other hand, in the local database $DB_3$, the level of substances `s1`, `s2`, and `s3` has exceeded once the related danger thresholds.

The following are examples of local transactions for $DB_1$:

1. $T =? \ sensor(s1) : t$ is a simple transaction that checks whether the substance `s1` has a presence over its danger threshold;
2. $T =? \ danger\_lv(X) : t; ? \ critical\_lv(Y) : t$ is a complex transaction determining for which substances $X$ the danger threshold has been exceeded and for which substances $Y$ it has been exceeded twice. $\diamond$

### *2.2 Amalgamated Active U-Datalog*

In order to amalgamate the static and dynamic knowledge deriving from the local databases, each represented by a AAU-Datalog program, a logical mediator, called *supervisor*, is used. In order to deal with the set of local databases, there is the need of univocally identifying each of them. To this purpose, we assume that:

- each local AAU-Datalog program is univocally identified by exactly one progressive numerical constant, starting from 1;
- the supervisor is univocally identified by letter "s".

The previous constants are called *NAME constants*. A *NAME-term* over a set of NAME-constants $C$ is either a subset of NAME-constants, or a variable ranging over NAME-constants (NAME-variable). In the following, in order to simplify notation, ground NAME-terms containing just one element will be denoted direcly by the element they contain. Thus, the NAME-term $\{1\}$ will be denoted by $1$.

An amalgamated atom (literal) is simply an annotated atom (literal) pointing out the identifier of the local database it belongs to. In particular, if $A : \psi$ is an annotated atom (literal) and $D$ is a NAME-term, then $A : [D, \psi]$ is an amalgamated atom (literal). An *Amalgamated Active U-Datalog* (or *AmAU-Datalog*) program is an AAU-Datalog program in which all the atoms are transformed into amalgamated atoms.

*Definition 8* (*Amalgamated atom, rule, and program*)
Let $\psi$ be an annotation over a $\mathcal{T}$-language, $D$ a NAME-term, and $A$ an atom (literal). Then, $A : [D, \psi]$ is an *amalgamated atom (literal)*. An Amalgamated Active U-Datalog rule (program) is an AAU-Datalog rule (program) in which each atom is replaced by an amalgamated atom. Given an AAU-Datalog atom, rule, or program $K$, we denote with $\mathbf{AT}(\mathbf{K})$ the corresponding amalgamated atom, rule, or program. In particular, if the AAU-Datalog program $DB_i$, having $i$ as identifier, contains atoms like $A : [\psi]$, $AT(DB_i)$ contains atoms like $A : [i, \psi]$. □

Based on the previous notion, the *supervisor database* can now be defined as a set of rules mediating the static and dynamic knowledge deriving from the local databases, hiding their internal structure. Thus, a supervisor is an amalgamated AAU-Datalog program, whose deductive rules refer to all the existing local AAU-Datalog local databases and whose active rules supply a simple and strong mechanism to exhibit a reaction in case of complex events involving several local databases of the amalgam. This is quite useful in supporting a global reaction in front of a situation that requires checking conditions concerning more that one local database. Differently from local databases, the supervisor is not characterized by a proper extensional database. Thus, no updates can be executed against it.

*Definition 9* (*Supervisor, strong supervisor*)
Let $DB_1, \ldots, DB_n$ be $n$ AAU-Datalog programs. A supervisor database $S$ is an AmAU-Datalog program composed of

- A set of amalgamated deductive rules $SIDB$ of the form

$$A : [s, \mu] \leftarrow A_1 : [D_1, \mu_1], \ldots, A_k : [D_k, \mu_k]|$$

  where each $D_i$, $i = 1, \ldots, k$, is a NAME-term over $\{s, 1, \ldots, n\}$.
- A set of amalgamated active rules $SAR$ of the form

$$\alpha_1 A_1 : [D_1, \mu_1], \ldots, \alpha_k A_k : [D_k, \mu_k] | L_{k+1} : [D_{k+1}, \mu_{k+1}], \ldots, L_n : [D_n, \mu_n] \to$$
$$\alpha_{n+1} A_{n+1} : [D_{n+1}, \mu_{n+1}], \ldots, \alpha_{n+m} A_{n+m} : [D_{n+m}, \mu_{n+m}]$$

  where $D_i$, $i = 1, \ldots, k, n+1, \ldots, n+m$ is a ground NAME-term containing just one element over $\{1, \ldots, n\}$ whereas $D_j$, $j = k+1, \ldots, n$, is a NAME-term over $\{s, 1, \ldots, n\}$.

A supervisor is *strong* if no amalgamated deductive rule contains literals like $L_j : [s, \mu_j]$ in its body.    □

A supervisor is *strong* if mediated knowledge does not depend on the supervisor itself. This notion will be used in providing a semantics for AmAU-Datalog programs. It is important to notice that the supervisor acts as a filter for the transactions to be executed against the local programs, as shown by the following example.

*Example 4*
Consider the case in which the supervisor does not contain any rule. All queries of type

$$? \, A : [s, \mu]$$

would return the unknown truth value ($\bot$), independently from the truth value assigned to $A$ in the single local AAU-Datalog programs.

If we want to assign the truth value $\mu$ to the ground atom $A$ only if each of the $n$ local programs assigns to $A$ the same value, the following amalgamated rule has to be inserted in the supervisor:

$$A : [s, V] \leftarrow A : [1, V], \ldots, A : [N, V]|$$

By using rules like the previous ones, it is possible to assign a greater priority to a given program, to a particular predicate, or to a particular fact. For example, in order to express the fact that program $i$ contains more significant information than the others concerning a given atom $A$, it is sufficient to insert in the supervisor the following amalgamated rule:

$$A[s, V] \leftarrow A(X) : [i, V]|$$

As another significant example, consider two local AAU-Datalog programs, say $DB_1$ and $DB_2$, respectively identified by 1 and 2, and FOUR as the source lattice of truth values. Suppose that:

- rule $p(X, Y) : t \leftarrow q(X, Y) : t| \, + m(X) : t$ is contained in $DB_1$, resulting in the following amalgamated rule:

$$p(X, Y) : [1, t] \leftarrow q(X, Y) : [1, t]| \, + m(X) : [1, t];$$

- rule $p(X, Y) : f \leftarrow q(X, Y) : f| \, - m(X) : t$ is contained in $DB_2$, resulting in the following amalgamated rule:

$$p(X, Y) : [2, f] \leftarrow q(X, Y) : [2, f]| \, - m(X) : [2, t].$$

If the two programs assign some incomparable truth values to a specific atom constructed over $p$, the supervisor can decide to assign to this atom a particular truth value. For example:

1) it may assign a combination of the values assigned to the atom by the local databases, through the amalgamated rule

$$p(X, Y) : [s, \sqcup\{Z, W\}] \leftarrow p(X, Y) : [1, Z], p(X, Y) : [2, W]|$$

2) or it may choose the value assigned by program 1, through the amalgamated rule

$$p(X, Y) : [s, Z] \leftarrow p(X, Y) : [1, Z], p(X, Y) : [2, W]| \, .    \diamond$$

By combining the knowledge represented by the local programs with the knowledge represented by the supervisor, we obtain the overall amalgam of static and dynamic knowledge. However, in order to complete the amalgamation process, there is the need of specifying how atoms like $A_k : [D_k, \mu_k]$ have to be defined, when $D_k$ is a ground NAME-term. Such an atom specifies that $A_k$ is true with truth value $\mu_k$ in all the databases whose identifiers are contained in $D_k$. The semantics for such atoms is provided by inserting in the amalgam an additional sets of rules, called *combination axioms*. Such rules specify that an atom $A_k : [D_k, \mu_k]$, $D_k = \{i_1, ..., i_n\}$, is true if the following conditions hold:

1. $A : [i_1, \mu_{i_1}],...,A : [i_n, \mu_{i_n}]$ are true;
2. $\mu_k = \sqcup\{\mu_{i_1}, ..., \mu_{i_n}\}$.

*Definition 10* (*Combination axioms*)
Let $D \subseteq \{1, \ldots, n\}$ be a set of NAME-constants, $A$ a deductive atom of the base language, and $\mu_D$ an annotation. Atom $A : [D, \mu_D]$ is defined by the following *combination axiom*

$$A : [D, \bigsqcup_{i \in D} \mu_i] \leftarrow \bigwedge_{i \in D} A : [i, \mu_i]$$

where $\mu_D = \bigsqcup_{i \in D} \mu_i$. In the following, combination axioms will be denoted by $C$. □

*Example 5*
Let $D = \{1, 2\}$. Consider the following amalgamated rules:

$C_1 = A : [1, \mu_0] \leftarrow A_1 : [1, \mu_1], \ldots, A_k : [1, \mu_k] |\ \alpha_{k+1} A_{k+1} : [1, \mu_{k+1}], \ldots, \alpha_n A_n[1, : \mu_n]$

$C_2 = A : [2, \overline{\mu}_0] \leftarrow A_1 : [2, \overline{\mu}_1], \ldots, A_l : [2, \overline{\mu}_l] |\ \alpha_{l+1} A_{l+1} : [2, \overline{\mu}_{l+1}], \ldots, \alpha_m A_m[2, : \overline{\mu}_m]$

The following is a combination axiom for atom $A$:

$A : [\{1, 2\}, \sqcup\{\mu_0, \overline{\mu}_0\}] \leftarrow A : [1, \mu_0], A : [2, \overline{\mu}_0] |\ .$ ◇

We are now able to define the *amalgam* of the local databases and the supervisor as a program combining the information represented by the various local databases, the supervisor, and the combination axioms.

*Definition 11* (*Amalgam*)
Let $DB_1, \ldots, DB_n$ be $n$ AAU-Datalog program, $DB_i = EDB_i \cup IDB_i \cup AR_i$, and $S = SIDB \cup SAR$ a supervisor. Let $C$ be the set of combination axioms. The *amalgam* $\mathcal{A}$ of $(S, DB_1, \ldots, DB_n)$ is the amalgamated program defined as follows:

$$\mathcal{A} = \cup_{i=1}^{n} AT(DB_i) \cup S \cup C.$$

The extensional part of the amalgam is $EDB_\mathcal{A} = \cup_{i=1}^{n}(AT(EDB_i))$, the intensional part is $IDB_\mathcal{A} = \cup_{i=1}^{n}(AT(IDB_i)) \cup SIDB \cup C$, and the active part is $AR_\mathcal{A} = \cup_{i=1}^{n}(AT(AR_i)) \cup SAR$. □

It is important to remark that, given a set of AAU-Datalog programs, the set of combination axioms that can be constructed is fixed, therefore they can be considered as implicitly defined by the amalgam semantics.

Similarly to AAU-Datalog programs, each operation that can be executed against an amalgam is called *transaction*. A transaction is a deductive rule without head and, as such, may execute a query, may require the execution of a set of updates against the local databases, and may trigger some active rules both at the local and global level. A transaction can be simple or complex, as pointed out by the following definition.

*Definition 12* (*Transaction*)
Let $DB_1, \ldots, DB_n$ be $n$ AAU-Datalog programs, $S$ a supervisor, $C$ a set of combination axioms and $\mathcal{A} = \cup_{i=1}^{n} AT(DB_i) \cup C \cup S$ the amalgam of $(S, DB_1, \ldots, DB_n)$. A *simple transaction* $T$ against $\mathcal{A}$ is an AmAU-Datalog deductive rule with no head and with a non-empty query part; a *complex transaction* is a sequence of simple transactions $T_1; \ldots; T_k$. □

```
IDB_S:  pblock(X):[s,V]←danger_lv(X):[1,V],danger_lv(X):[2,V],danger_lv(X):[3,V]|
                     +partial_block(X):[1,V],+partial_block(X):[2,V],
                     +partial_block(X):[3,V].
        pblock(X):[s,⊥]←danger_lv(X):[{1,2,3},⊥]| +reread(X):[1,t],+reread(X):[2,t],
                     +reread(X):[3,t].
        pblock(X):[s,⊥]←danger_lv(X):[{1,2,3},⊤]| +reread(X):[1,t],+reread(X):[2,t],
                     +reread(X):[3,t],+partial_block(X):[1,⊥],
                     +partial_block(X):[2,⊥],+partial_block(X):[3,⊥].
        tblock(X,Y):[s,t]←partial_block(X):[W,t],partial_block(Y):[Z,t],not_eq(X,Y):t|
                     +total_block(X,Y):[1,t],+total_block(X,Y):[2,t],
                     +total_block(X,Y):[3,t].
        tblock(X,Y):[s,t]←critical_lv(X):[W,t],critical_lv(Y):[Z,t],not_eq(X,Y):t|
                     +total_block(X,Y):[1,t],+total_block(X,Y):[2,t],
                     +total_block(X,Y):[3,t].
AR_S:   +partial_block(X):[Y,V]| total_block(W,Z):[Y,t]→-partial_block(X):[Y,V].
        +total_block(X,Y):[Z,t]| partial_block(W):[Z,V]→-partial_block(W):[Z,V].
        +reread(X):[Y,t]| partial_block(X):[H,t]→-partial_block(X):[H,t].
        +reread(X):[Y,t]| total_block(X,Z):[H,t]→-total_block(X,Z):[H,t].
        +reread(X):[Y,t] | total_block(Z,X):[H,t]→-total_block(Z,X):[H,t].
```

Figure 2. The supervisor database.

*Example 6*
Consider Example 3. In order to decide partial and total blocks of car circulation, suppose that the following policy is applied. Car circulation must be partially forbidden if all the sensors determine that a monitored substance has a presence above its danger threshold. On the other hand, car circulation is totally forbidden if one of the following conditions holds:

- all the sensors determine that two different substances have a presence above their danger threshold;
- for two different substances, a critical situation is detected by some sensors.

The detection of partial and total blocks requires a cooperation among the various sensors, therefore this functionality can be assigned to an AmAU-Datalog supervisor database. Based on the previous policy, such database can be designed as illustrated in Figure 2.

Predicate `pblock` manages partial blocks. In particular, if all the sensors agree in assigning the truth value to atom `sensor(X)`, it assigns the same truth value to the atom concerning partial block of car circulation. On the other hand, if all the sensors agree in assigning a truth value $\bot$ or $\top$ to a given substance, some facts are asserted specifying

```
EDB_A: sensor(s1):[1,t],sensor(s2):[1,⊥],sensor(s3):[1,t],over_lv(s1):[1,t],
       critical_lv(s1):[1,t],over_lv(s3):[1,t],partial_block(s3):[1,t],
       sensor(s1):[2,⊥],sensor(s2):[2,t],sensor(s3):[2,t],over_lv(s2):[2,t],
       critical_lv(s2):[2,t],over_lv(s3):[2,t],partial_block(s3):[2,t],
       sensor(s1):[3,t],sensor(s2):[3,t],sensor(s3):[3,t],over_lv(s1):[3,t],
       over_lv(s2):[3,t],over_lv(s3):[3,t],partial_block(s3):[3,t].
IDB_A: danger_lv(X):[i,Y]←sensor(X):[i,Y] | i=1,2,3.
       danger_lv(X):[i,t]←sensor(X):[i,t] | +over_lv(X):[i,t], i=1,2,3.
       danger_lv(X):[i,t]←sensor(X):[i,t],over_lv(X):[i,t]| +critical_lv(X):[1,t],
                                                                           i=1,2,3.
       danger_lv(X):[i,f]←sensor(X):[i,f],over_lv(X):[i,t] | -over_lv(X):[i,t],
                                                                           i=1,2,3.
       pblock(X):[s,V]←danger_lv(X):[1,V],danger_lv(X):[2,V],danger_lv(X):[3,V] |
                       +partial_block(X):[1,V],+partial_block(X):[2,V],
                       +partial_block(X):[3,V].
       pblock(X):[s,⊥]←danger_lv(X):[{1,2,3},⊥] | +reread(X):[1,t],+reread(X):[2,t],
                       +reread(X):[3,t].
       pblock(X):[s,⊥]←danger_lv(X):[{1,2,3},⊤]| +reread(X):[1,t],+reread(X):[2,t],
                       +reread(X):[3,t],+partial_block(X):[1,⊥],
                       +partial_block(X):[2,⊥],+partial_block(X):[3,⊥].
       tblock(X,Y):[s,t]←partial_block(X):[W,t],partial_block(Y):[Z,t],not_eq(X,Y):t|
                       +total_block(X,Y):[1,t],+total_block(X,Y):[2,t],
                       +total_block(X,Y):[3,t].
       tblock(X,Y):[s,t]←critical_lv(X):[W,t],critical_lv(Y):[Z,t],not_eq(X,Y):t|
                       +total_block(X,Y):[1,t],+total_block(X,Y):[2,t],
                       +total_block(X,Y):[3,t].
       sensor(s1):[{1,2},⊔{V_1,V_2}]←sensor(s1):[1,V_1],sensor(s1):[2,V_2]| .
       sensor(s1):[{1,3},⊔{V_1,V_3}]←sensor(s1):[1,V_1],sensor(s1):[3,V_3]| .
       sensor(s1):[{2,3},⊔{V_2,V_3}]←sensor(s1):[2,V_2],sensor(s1):[3,V_3]| .
       sensor(s1):[{1,2,3},⊔{V_1,V_2,V_3}]←sensor(s1):[1,V_1],sensor(s1):[2,V_2],
                                           sensor(s1):[3,V_3]| .
       ...      ...      ...
AR_A:  -over_lv(X):[i,t]| critical_lv(X):[i,t]→-critical_lv(X):[i,t], i=1,2,3.
       +partial_block(X):[Y,V]| total_block(W,Z):[Y,t]→-partial_block(X):[Y,V].
       +total_block(X,Y):[Z,t]| partial_block(W):[Z,V]→-partial_block(W):[Z,V].
       +reread(X):[Y,t]| partial_block(X):[H,t]→-partial_block(X):[H,t].
       +reread(X):[Y,t]| total_block(X,Z):[H,t]→-total_block(X,Z):[H,t].
       +reread(X):[Y,t]| total_block(Z,X):[H,t]→-total_block(Z,X):[H,t].
```

Figure 3. The amalgam of $DB_1, DB_2, DB_3$ and $S$.

that the level for that substance has to be read again. In the first case, this is because local databases have no information concerning the level of that substance, in the second case because local databases do not agree on such level. Note that in the first case, updates are indirectly performed by the first rule. Predicate tblock manages total blocks, as specified before. Since a total block is always due to two substances, predicate total_block specifies also their names. The supervisory database also contains five active rules. In particular:

- the first active rule requires the deletion of a just inserted partial block information if a total block information has already been detected. In this case, partial block information is discarded by the evaluation process;
- the second rule deletes all partial blocks information when a total block is detected;
- the last three rules delete all partial or total blocks information concerning a given substance $X$ for which the danger level has to be re-read.

Figure 3 illustrates the amalgam $\mathcal{A}$ of the supervisor together with the local databases of the Example 3. In Figure 3 only some few combination axioms of $\mathcal{A}$ are presented; all the others axioms are computed as described in Definition 10.

The following are examples of transactions for $\mathcal{A}$:

1. $T =?\ pblock(X) : [s, t]$ is a simple transaction checking which substances generate a partial block;
2. $T =?\ pblock(X) : [s, V];\ ?\ tblock(Y, Z)[s, t]$ is a complex transaction determining the substances generating partial and total blocks.                                    $\diamond$

## 3 Amalgamated Active-U-Datalog: the deductive semantics

In defining the semantics of an AmAU-Datalog program, we consider as observable properties of the transaction execution: the set of bindings (the answer), the new database state, and the indication of success (commit/abort) of the transaction itself. The semantics of AmAU-Datalog programs, with respect to a simple transaction, can be defined in three steps:

- In the first step, the semantics of the amalgam is computed, collecting the set of bindings that satisfy the query and the requested updates. In this step, no update is executed and only deductive reasoning is performed. All the updates generated during this phase are gathered together but not executed.
- In the second step, the semantics of the active part of the program is computed, according to the model and the updates collected in the first step. The result of this step is the set of updates generated either by the deductive and/or the active part of the amalgam. In this step, conflicting updates have to be removed. This is possible by applying a *parametric conflict resolution policy*.
- In the third step, the updates obtained from the second step are executed against the extensional database.

The previous three steps are repeated for each simple transaction contained in a complex transaction. Hence, the state of the database evolves after each simple transaction execution.

In the following, we describe in details the first semantic step. The other steps will be formally introduced in Sections 4 and 5.

In order to compute the semantics of the deductive part of the amalgam, we first specify how the deductive semantics of the local databases is computed and then we describe how the local semantics are combined together to form the semantics of the amalgam. In the following, the term "rule" refers to a deductive rule.

### 3.1 Deductive semantics of an AAU-Datalog program

In the following we assume that $\mathcal{T}$ is a complete lattice containing the least element $\bot$. In order to define the semantics of the deductive part of an AAU-Datalog program, we consider an *Extended Herbrand Base* defined as follows.

*Definition 13* (*Extended Herbrand base*)
The Extended Herbrand Base $\overline{\mathcal{CB}}_\mathcal{L}$ of a base language $\mathcal{L}=(\Sigma, \Pi, V)$ is composed of *constrained* ground atoms of the form $A \leftarrow \alpha_1 D_1, \ldots, \alpha_k D_k$, where $A$ is a ground ($\Pi^i \cup \Pi^e, \Sigma, V$)-atom and $\alpha_j D_j, j = 1, \ldots, k$, are ground ($\Pi^u, \Sigma, V$)-atoms.                                                                      $\square$

Since the language does not contain function symbols, the Extended Herbrand Base is a finite set. Similarly to standard logic programming, the Extended Herbrand Base is then used to define interpretations. However, differently from the standard framework, an interpretation must assign to each constrained atom, belonging to the Extended Herbrand Base, a truth value, constructed starting from the considered lattice. Such truth values can be formalized by introducing the concept of *ideal* and of *principal*.

*Definition 14* (*Ideal (Kifer and Subrahmanian, 1992)*)
An ideal of an upper semilattice[5] $\mathcal{T}$ is any subset $K$ of $\mathcal{T}$ such that:

1. $K$ is downward closed, that is $s \in K, t \leq s \Rightarrow t \in K$;
2. $K$ is closed with respect to the finite least upper bound operation, that is $s, t \in K \Rightarrow s \sqcup t \in K$.

An ideal $K$ is *principal* if for some $p \in \mathcal{T}$, $K = \{s \mid s \leq p\}$. In this case, we use $p$ to denote $K$. An ideal $I$ is null if $I = \{\}$. A principal $P$ is null if $P = \bot$. We let $\sqcup\{\} = \bot$. □

Now suppose to define an interpretation as a function assigning an ideal (principal) to each constrained atom. This means that the truth value is assigned to the constrained atom as a whole and not to the various atoms composing the constrained atom, as we would like to do.

In order to overcome the previous limitation, the intepretation is defined as a function that, given a constrained atom, assigns a truth value to each atom appearing in it. Thus, if the constrained atom contains $n - 1$ update atoms in its body, the interpretation function must assign to the constrained atom $n$ truth values, one for the head of the constrained atom and $n - 1$ for the update atoms in the body. In order to formalize this concept, it is useful to introduce the concept of $n$-ideal and $n$-principal.

*Definition 15* ($n$-*ideal*, $n$-*principal*)
Let $\mathcal{T}$ an upper semilattice. An $n$-ideal of an upper semilattice $\mathcal{T}$ is a tuple composed of $n$ ideals of $\mathcal{T}$; with $n$-I we denote the set of all the $n$-ideals of $\mathcal{T}$. An $n$-principal of an upper semilattice $\mathcal{T}$ is a tuple composed of $n$ principals of $\mathcal{T}$; with $n$-P we denote the set of all the $n$-principals of $\mathcal{T}$. An $n$-ideal is null (denoted by $\delta$) if all components represent the null ideal. An $n$-principal is null (denoted by $\delta_r$) if all components represent the null principal. Moreover, the following sets are introduced: $N\text{-}I = \bigcup_n n\text{-}I$ and $N\text{-}P = \bigcup_n n\text{-}P$. We let $\sqcup \delta = \delta_r$. □

Note that, since a principal is also an ideal, an $n$-principal is an $n$-ideal. The main difference is that each single element of an $n$-principal is an ideal that can be identified by a single value whereas each single element of an $n$-ideal is a generic ideal.

It is also important to note that $n$-ideals ($n$-principals) of a complete semilattice $\mathcal{T}$ can be seen as ideals (principals) of the complete lattice $\mathcal{T}^n$, in which the order $\leq$ is defined component by component, in the obvious way (Subrahmanian, 1994).

Intepretations can now be defined as follows.

---

[5] A set $R$ is an upper semilattice, with respect to a given ordering, if any pair of elements of $R$ has a least upper bound in $R$.

*Definition 16 (Interpretations)*
Let $\mathcal{L}$ be the base language and $\mathcal{T}$ a complete lattice. An *Herbrand interpretation* (*restricted Herbrand interpretation* - r-interpretation-) $I$ of $\mathcal{L}$ is any total map from the extended Herbrand Base $\overline{\mathcal{CB}}_{\mathcal{L}}$ to $N$-$I$ ($N$-$P$). Thus, each constrained atom is mapped into an $n$-ideal ($n$-principal). We assume that the first ideal of the $n$-ideal is assigned to the head of the constrained atom whereas all the other ideals are assigned to update atoms in the body, following the ordering by which they appear.  □

Based on the previous definition, an interpretation can always be seen as a set of ground and c-annotated constrained atoms and in the following we often use this notation. Suppose $I(A \leftarrow \alpha_1 D_1, \ldots, \alpha_k D_k) = <\mu_0, \mu_1, \ldots, \mu_k>$. In the following we denote $\mu_0$ with $I_0(A \leftarrow \alpha_1 D_1, \ldots, \alpha_k D_k)$ and $\mu_j$ with $I_j(A \leftarrow \alpha_1 D_1, \ldots, \alpha_k D_k)$, $j = 1, \ldots, k$. Moreover, to simplify the notation, we assume that all constrained atoms mapped to the null $n$-ideal are not specified when describing the interpretation and all constrained atoms mapped to the null $n$-principal are not specified when describing the r-interpretation.

When $\mathcal{T}$ is a complete lattice (according to the order $\leq$), the order $\leq$ can be extended point to point to the r-interpretations as follows: let $I, \overline{I}$ be r-interpretations, then $I \leq \overline{I}$ if and only if, for each $A \leftarrow \alpha_1 D_1, \ldots, \alpha_k D_k \in \overline{\mathcal{CB}}_{\mathcal{L}}$, the following condition holds:

$$I_j(A \leftarrow \alpha_1 D_1, \ldots, \alpha_k D_k) \leq \overline{I}_j(A \leftarrow \alpha_1 D_1, \ldots, \alpha_k D_k), \quad j = 0, \ldots, k.$$

Starting from the notion of r-interpretation, it is possible to define when a constrained ground atom is true in an interpretation.

*Definition 17 (R-satisfaction)*
Let $I$ be an interpretation, $<\mu_0, \mu_1, \ldots, \mu_k>$ a list of c-annotations on a lattice $\mathcal{T}$, $L \equiv A \leftarrow \alpha_1 D_1, \ldots, \alpha_k D_k \in \overline{\mathcal{CB}}_{\mathcal{L}}$. $I$ satisfies the ground constrained annotated atom $M \equiv A : \mu_0 \leftarrow \alpha_1 D_1 : \mu_1, \ldots, \alpha_k D_k : \mu_k$ ($I \models M$) if and only if $I(L) \subseteq <\mu_0, \mu_1, \ldots, \mu_k>$. Let $I$ be an r-interpretation, $<\mu_0, \mu_1, \ldots, \mu_k>$ a list of c-annotations on a lattice $\mathcal{T}$, $L \equiv A \leftarrow \alpha_1 D_1, \ldots, \alpha_k D_k \in \overline{\mathcal{CB}}_{\mathcal{L}}$. $I$ r-satisfies the ground constrained annotated atom $M \equiv A : \mu_0 \leftarrow \alpha_1 D_1 : \mu_1, \ldots, \alpha_k D_k : \mu_k$ ($I \models^r M$) if and only if $\mu_0 \leq I_0(C)$ and $\mu_j \leq I_j(L), j = 1, \ldots, k$.  □

*Example 7*
Let $\mathcal{T}$ be $\mathcal{R}[0,1]$, $A \leftarrow +D_1, +D_2 \in \overline{\mathcal{CB}}_{\mathcal{L}}$. Let $I$ be an r-interpretation such that

- $I(A \leftarrow +D_1, +D_2) = (0.22, 0.5, 0.6)$.
- for any other constrained ground atom $K \in \overline{\mathcal{CB}}_{\mathcal{L}}$, $I(K) = \delta_r$.

$I$ satisfies the ground constrained annotated atom $A : 0.2 \leftarrow +D_1 : 0.3, +D_2 : 0.5$ since $0.2 < 0.22$, $0.3 < 0.5$, and $0.5 < 0.6$.  ◇

Based on the previous concepts, we can now define a fixpoint operator $T_P$, computing the semantics of the deductive part. Intuitively, given an interpretation $I$, operator $T_P(I)$ applied to a constrained ground atom $A$ returns the least n-ideal for $A$, containing all the $n$-ideals for $A$ that can be computed starting from rules in $GI(IDB_P \cup EDB_P)$,[6] whose body is satisfied by $I$. The operator introduced in the following definition extends the one presented in (Kifer and Subrahmanian, 1992) to deal with constrained atoms and $n$-ideals. In particular, the computation gathers updates generated from atoms in the body of a rule and assign them to the atom in the head.

---

[6] We recall that $GI(P)$ represents all the c-annotated, ground instances of $P$.

*Definition 18* (*Operator $T_P$*)
Let $\mathcal{T}$ be a complete lattice, $I$ an interpretation and $P = IDB_P \cup EDB_P \cup AR_P$ an AAU-Datalog program. For any constrained ground atom $A \in \overline{\mathcal{CB}_\mathcal{L}}$, $A \equiv H \leftarrow \alpha' D'_1, \ldots, \alpha'_k D'_k$, $T_P(I)(A)$ is defined as the least $n$-ideal of $\mathcal{T}$ containing the following set:

$$\{<\mu'_0, \mu'_1, \ldots, \mu'_k>|$$
$$\exists H : \mu_0 \leftarrow B_1 : \mu_1, \ldots, B_n : \mu_n|\ \alpha_{n+1} D_{n+1} : \mu_{n+1}, \ldots,$$
$$\alpha_{n+p} D_{n+p} : \mu_{n+p} \in GI(IDB_P \cup EDB_P)$$
$$\exists r_1, \ldots, r_n \quad r_u = \overline{B}_u : \mu_u \leftarrow \alpha_u^1 D_u^1 :, \ldots, \alpha_u^{n_u} D_u^{n_u}, \ u = 1, \ldots, n$$
$$I \models r_u, \ u = 1, \ldots, n$$
$$\{\alpha'_1 D'_1 : \mu'_1, \ldots, \alpha'_k D'_k : \mu'_k\} =$$
$$= \{\alpha_{n+1} D_{n+1} : \mu_{n+1}, \ldots, \alpha_{n+p} D_{n+p} : \mu_{n+p}\} \stackrel{+}{\cup} \left( \bigcup_{\substack{u=1,\ldots,n \\ t=1,\ldots,n_u}}^{+} \{\alpha_u^t D_u^t : \mu_u^t\} \right) \}$$

The set $A \stackrel{+}{\cup} B$ is defined as the usual set union with the following exception: if $A$ contains $\alpha D : \mu_1$ and $B$ contains $\alpha D : \mu_2$, then $A \stackrel{+}{\cup} B$ contains $\alpha D : \sqcup\{\mu_1, \mu_2\}$ and does not contain $\alpha D : \mu_1$ and $\alpha D : \mu_2$. □

*Example 8*
Let $T = (\mathcal{C}, \mathcal{F}, \mathcal{V})$ and $(\Pi, \Sigma, V)$ be respectively the $\mathcal{T}$-language and the *base language* of Example 2 and $P$ the AAU-Datalog program defined as follows:

$IDB_P$: 
```
r1: q1(a,b):0.5←p1(a,b):0.5| +p2(a,b):0.75.
r2: q1(Y1,Y2):X1←p1(Y1,Y2):X1,q2(Y1,Y2):X1| +p2(Y1,Y2):X1.
r3: q2(Y3,Y4):X2←p3(Y3,Y4):X2,p3(Y4,Y5):X3| +p2(Y3,Y4):X3,
    +p3(Y3,Y4):X3.
```
$AR_1$: `ar1: +p2(a,b):0.75 | p1(a,b):0.5→+p1(a,b):0.75`

$EDB_P$: `p1(a,b):0.5,p3(a,b):0.5,p3(b,c):0.75.`

Let $A \in \overline{\mathcal{CB}_\mathcal{L}}$, $A \equiv q_1(a,b) \leftarrow +p_2(a,b), +p_3(a,b)$, and $I$ be the following interpretation (which is also an r-interpretation):

$$\begin{aligned}
I(p_1(a,b)) &= 0.5 \\
I(p_3(a,b)) &= 0.5 \\
I(p_3(b,c)) &= 0.75 \\
I(q_1(a,b) \leftarrow +p_2(a,b)) &= (0.5, 0.75) \\
I(q_2(a,b) \leftarrow +p_2(a,b), +p_3(a,b)) &= (0.5, 0.75, 0.75) \\
I(K) &= \delta_r \text{ for any other constrained ground} \\
& \quad \text{atom } K \in \overline{\mathcal{CB}_\mathcal{L}}
\end{aligned}$$

$T_P(I)(A)$ is the $n$-principal $(0.5, 0.75, 0.75)$. Indeed:

- $q_1(a,b) : 0.5 \leftarrow p_1(a,b) : 0.5, q_2(a,b) : 0.5| + p_2(a,b) : 0.5 \in GI(IDB_P \cup EDB_P)$;
- $p_1(a,b) : 0.5 \in I, p_3(a,b) : 0.5 \in I, p_3(b,c) : 0.75 \in I, q_2(a,b) : 0.5 \leftarrow +p_2(a,b) : 0.75 \in I, +p_3(a,b) : 0.75 \in I$;
- $\{+p_2(a,b) : 0.5)\} \stackrel{+}{\bigcup} \{+p_2(a,b) : 0.75, +p_3(a,b) : 0.75)\} = \{+p_2(a,b) : \sqcup_2\{0.5, 0.75\}, +p_3(a,b) : 0.75\}$, thus obtaining the set $\{+p_2(a,b) : 0.75, +p_3(a,b) : 0.75\}$.   ◇

Given a constrained atom $A$ and an interpretation $I$, $T_P(I)(A)$ is an $n$-ideal, thus it is a set of tuples of lattice values.[7] However, $n$-principals more clearly characterize the semantics value of a given constrained atom, since they can be represented by using just one tuple of lattice value. Thus, they are good candidates for defining a bottom-up semantics for AAU-Datalog databases.

In order to use $n$-principals in defining the semantics of a AAU-Datalog program, a new fixpoint operator has to be defined dealing with $n$-principals. Similarly to what has been done in (Kifer and Subrahmanian, 1992), such an operator can be easily defined starting from operator $T_P$, by combining together all the tuples of ideals generated for a given atom. More formally, the operator, denoted by $R_P$, can be defined as follows.

*Definition 19 (Operator $R_P$)*

Let $I$ be an r-interpretation and $P$ an AAU-Datalog program. The operator $R_P(I)$, for any $A \in \overline{\mathcal{CB}_\mathcal{L}}$, is such that $R_P(I)(A) = \sqcup T_P(I)(A)$, where $\sqcup T_P(I)(A)$ is a shorthand for $\sqcup_{\tilde{t} \in T_P(I)(A)} \tilde{t}$.   □

$T_P$ and $R_P$ satisfy several important properties, as pointed out by the following theorem.

*Theorem 1*

Let $P$ be an AAU-Datalog program, $I$ an r-interpretation. Then:

1. $T_P$ is monotonic and continuous;
2. $R_P$ is monotonic;
3. $T_P \uparrow \omega = lfp(T_P)$;[8]
4. $lfp(T_P)$ is a model for $P$.

**Proof:** It follows from (Kifer and Subrahmanian, 1992), by considering $\mathcal{T}^n$ as a complete lattice.   □

Since we would like to use $R_P$ to compute the semantics of a AAU-Datalog program, thus assigning a single $n$-principal to each constrained atom, we must be able to compute the fixed point of $R_P$. The main difference between $T_P$ and $R_P$ is that the first is continuous while the second is not, as shown by the following example, taken from (Kifer and Subrahmanian, 1992). This implies that $R_P \uparrow \omega$ is not a fixed point for $R_P$.

---

[7] Note that in the previous example we deal with $n$-principals, but in general the $T_P$ operator generates an $n$-ideal.
[8] As usual, $T_P^1(I) = T_P(I)$, $T_P^{i+1} = T_P(T_P^i(I))$, and $T_P \uparrow \omega = \bigcup_i T_P^i(\{\})$ and $T_P \uparrow \omega(A) = \bigcup_i T_P^i(\{\})(A)$.

*Example 9*
(Kifer and Subrahmanian, 1992) Suppose $\mathcal{T}$ is the interval of real numbers $[0,1]$ with the usual ordering and consider the following program:
$p:0 \leftarrow |$
$p: \frac{1+x}{2} \leftarrow p:x|$
$q:1 \leftarrow p:1|$ .

It is quite easy to prove that $T_P \uparrow i$, $0 \leq i \leq \omega$, always assigns the empty ideal $\{\}$ to $q$. Therefore, $T_P \uparrow \omega(q) = \{\}$. According to the restricted semantics $R_P \uparrow \omega(p) = \{a|a \leq 1\}$. On the other hand, according to the general semantics, $T_P \uparrow \omega(p) = \{a|a < 1\}$. This means that from the r-semantics we obtain $R_P(R_P \uparrow \omega)(q) = \mathcal{T}$ while from the general semantics we obtain $T_P(T_P \uparrow \omega)(q) = \{\}$. Thus, $R_P \uparrow \omega$ is not a fixpoint of $R_P$ in the r-semantics but, in the general semantics $T_P \uparrow \omega$ is a fixpoint of $T_P$. $\diamond$

In order to make valid the equation $R_P \uparrow \omega = lfp(R_P)$ and using $R_P$ for computing the deductive semantics of an AAU-Datalog program, some sufficient conditions have to be proposed. To this purpose, in (Kifer and Subrahmanian, 1992) the notion of *acceptable program* has been introduced. In the following, the notion of acceptable program is revised to deal with constrained atoms. As we will see, for any acceptable program $P$, $R_P \uparrow \omega = lfp(R_P)$. Informally, an acceptable program $P$ is a program in which any c-annotated literal appearing in the body of a rule in $P$ is reachable by $T_P$ in at most $\omega$ step.

*Definition 20* (*Acceptable program*)
An AAU-Datalog program $P$ is said to be *acceptable* if and only if for any c-annotated literal $A:\mu$ appearing in the body of a rule $r$ of $P$, for any constrained annotated atom $M \in \overline{\mathcal{CB}}_\mathcal{L}$ with the same head, if $\sqcup(T_P \uparrow \omega) \models M'$, for some ground instance $M'$ of $M$, then $T_P \uparrow \omega \models M'$. In the previous formula, $\sqcup(T_P \uparrow \omega)$ is a shorthand for $\bigcup_A \sqcup T_P \uparrow \omega(A)$. $\square$

The following proposition presents a syntactic characterization of acceptable programs.

*Proposition 1*
Let $P$ be an AAU-Datalog program such that one of the following conditions holds:

1. all rules contain only c-annotated atoms in their bodies;
2. all rules contain only v-annotated atoms in their bodies.

Then $P$ is acceptable.

**Proof Sketch:**

1. Under the hypothesis, $T_P \uparrow \omega(A)$ is a finitely generated ideal for every $A$ and it is also an $n$-principal. Thus, $\sqcup(T_P \uparrow \omega) = T_P \uparrow \omega$ and therefore the thesis follows.
2. The thesis trivially follows because, under the hypothesis, no condition has to be checked. $\square$

If a program $P$ is acceptable then $R_P \uparrow \omega = lfp(R_P)$ holds and is possible to establish a relationship between $R_P$ and $T_P$.

*Theorem 2*
Let $P$ be an acceptable AAU-Datalog program, $\mathcal{T}$ a complete lattice. Suppose that $\mathcal{T}$ is endowed with the least element $\bot$, then $R_P \uparrow \omega = lfp(R_P) = \sqcup(lfp(T_P))$. The *fixpoint semantics* $\mathcal{F}$ of $P$ is defined as $\mathcal{F}(P) = lfp(R_P)$. □

In order to prove the previous theorem, we need a lemma.

*Lemma 1*
Let $P$ be an acceptable AAU-Datalog program over a complete upper semilattice $\mathcal{T}$ with the least element $\bot$. Then:

1. $R_P \uparrow \omega = \sqcup T_P \uparrow \omega$.
2. $R_P(R_P \uparrow \omega) = R_P \uparrow \omega$.

*Proof*
1. In order to prove the first statement we prove that $R_P \uparrow \omega \supseteq \sqcup(T_P \uparrow \omega)$ and viceversa.

   (a) $R_P \uparrow \omega \supseteq \sqcup(T_P \uparrow \omega)$: The proof of this case follows from the proof presented in (Kifer and Subrahmanian, 1992), by considering $\mathcal{T}^n$ as complete lattice. The proof does not use the notion of acceptability (which has been changed with respect to (Kifer and Subrahmanian, 1992)) but only the properties of $T_P$ and $R_P$, presented in Theorem 1.

   (b) $R_P \uparrow \omega \subseteq \sqcup(T_P \uparrow \omega)$: We show that, for all $i$ and $A \in \overline{\mathcal{CB}_\mathcal{L}}$

   $$\sqcup(T_P \uparrow \omega)(A) \supseteq (R_P \uparrow i)(A).$$

   From the previous statement, it follows that:
   $\sqcup(T_P \uparrow \omega)(A) \supseteq \{(R_P \uparrow i)(A) | i = 1, 2, ....\} = R_P \uparrow \omega$
   which is the thesis. We prove (1b) by induction on $i$.
   The base case is trivial since $\delta_r = \sqcup \delta$, by definition. For the inductive step, we assume that $\sqcup(T_P \uparrow \omega)(A) \supseteq (R_P \uparrow k)(A)$ and we prove that it holds also for $k+1$.
   By definition of $R_P$,
   $R_P(R_P \uparrow k)(A) = \sqcup \{< \mu'_0, \mu'_1, ..., \mu'_k > | H : \mu_0 \leftarrow B_1 : \mu_1, \ldots, B_n : \mu_n |$
   $\quad \alpha_{n+1} D_{n+1} : \mu_{n+1}, \ldots, \alpha_{n+p} D_{n+p} : \mu_{n+p} \in GI(IDB_P \cup EDB_P),$
   $\exists r_1, ..., r_n, r_u \equiv \overline{B}_u : \mu_u \leftarrow \alpha^1_u D^1_u : \mu^1_u, \ldots, \alpha^{n_u}_u D^{n_u}_u : \mu^{n_u}_u, u = 1, \ldots, n$
   $R_P \uparrow k \models^r r_i, i = 1, \ldots, n$
   $A \equiv H \leftarrow \alpha'_1 D'_1, \ldots, \alpha'_k D'_k$

   $\{\alpha'_1 D'_1 : \mu'_1, \ldots, \alpha'_k D'_k : \mu'_k\} =$

   $= \{\alpha_{n+1} D_{n+1} : \mu_{n+1}, ..., \alpha_{n+p} D_{n+p} : \mu_{n+p}\} \overset{+}{\bigcup} \left( \underset{\substack{u=1,\ldots,n \\ t=1,\ldots,n_u}}{\overset{+}{\bigcup}} \{\alpha^t_u D^t_u : \mu^t_u\} \right) \}$

   By the inductive assumption, we have $\sqcup(T_P \uparrow \omega)(A) \supseteq (R_P \uparrow k)(A)$ and therefore

   $$\sqcup(T_P \uparrow \omega) \models \overline{B}_u : \mu_u \leftarrow \alpha^1_u D^1_u : \mu^1_u, \ldots, \alpha^{n_u}_u D^{n_u}_u : \mu^{n_u}_u, u = 1 \ldots, n$$

Because of the acceptability of $P$, this also means that

$$T_P \uparrow \omega \models \overline{B}_u : \mu_u \leftarrow \alpha_u^1 D_u^1 : \mu_u^1, \ldots, \alpha_u^{n_u} D_u^{n_u} : \mu_u^{n_u}, \ u = 1 \ldots, n$$

and therefore $< \mu'_0, \mu'_1, ..., \mu'_k > \in T_P \uparrow \omega$. From this, the thesis follows.

2. The proof of the second statement follows from the monotonicity of $R_P$ and since $R_P \uparrow \omega \supseteq R_P \uparrow i$, for all $i$. The other inclusion is proved similarly to the proof of the first statement, case 1b $\square$

From the proof of Lemma 1 it is now easy to prove Theorem 2.

**Proof of Theorem 2**. By statement (1) of Lemma 1 and by Theorem 1, it follows that $R_P \uparrow \omega = \sqcup(lfp(T_P))$. By (2) of Lemma 1, $R_P \uparrow \omega$ is a fixpoint of $R_P$, and, due to the monotonicity of $R_P$, it is the least fixpoint. $\square$

### *3.2 Deductive semantics of the amalgam*

Starting from the deductive semantics of the local databases, it is quite easy to compute the deductive semantics of an AmAU-Datalog program. Indeed, all the presented definitions and results still hold when applied to amalgamated atoms, since such atoms only differ from the annotated ones due to the presence of a label indicating the database to which they refer. Thus, the semantics presented for AAU-Datalog programs can also be applied to AmAU-Datalog programs.

In the following, to emphasize the fact that we are considering the fixpoint operator associated with an amalgam, an r-interpretation for an amalgam is called *AM-interpretation* and operator $R_P$ is renamed as $AM_P$.

It is now interesting to establish some connections between the semantics of the amalgam and the semantics of the local databases. To this purpose, similarly to what has been done in (Subrahmanian, 1994), we introduce two notions that allow one to relate r-interpretations and AM-interpretations. In particular, given an r-interpretation, we call *locale* the set of AM-interpretations that agree with that r-interpretation as far as a particular local database is concerned. On the other hand, given an AM-interpretation I, the *projection* of $I$ on a given local database $DB$ is the (unique) r-interpretation obtained by restricting I to $DB$. In the following, given an AM-interpretation $I$, we denote with $I(B)(i)$ the $n$-principal assigned to $B$ in $DB_i$.

*Definition 21* (*Locale*)
Let $I^{loc}$ be an r-interpretation for an AAU-Datalog program $DB_i$. The *locale of* $I^{loc}$ is the set $\{I^{aml} \ | \ I^{aml} \ is \ $ an AM-interpretation and $\forall$ constrained ground atom $B$, $I^{aml}(B)(i) = I^{loc}(B)\}$. $\square$

*Definition 22* (*Projection*)
Let $\mathcal{A}$ be the amalgam of $(S, DB_1, \ldots, DB_n)$ and $I^{aml}$ an AM-interpretation. The *projection of $I^{aml}$ on $DB_i$*, with $1 \leq i \leq n$, is the r-interpretation $I^{loc}$ defined as $I^{loc}(A) = I^{aml}(A)(i)$. $\square$

*Example 10*
Let $DB_1, DB_2$ be two local AAU-Datalog databases. Suppose that the base language contains only the extensional unary predicates $p$ and $q$. Let FOUR be the lattice of truth values, and $I^{aml}$ the following A-interpretation:

$I^{aml}(p(a))(1) = t.$      $I^{aml}(p(a))(2) = f.$
$I^{aml}(q(a))(1) = \top.$      $I^{aml}(q(a))(2) = t.$

The projection $I^{loc}$ of $I^{aml}$ on $DB_1$ is the following r-interpretation $I$:

$I^{loc}(p(a)) = t.$      $I^{loc}(q(a)) = \top.$      ◇

Starting from the previous notions, the following results can be stated, pointing out the relationships between the fixed point semantics of the amalgam and that of the local databases.

*Theorem 3*
Let $DB_1, \ldots, DB_n$ be $n$ AAU-Datalog programs, $S$ a supervisor, and $\mathcal{A}$ the amalgam of $(S, DB_1, \ldots, DB_n)$. Let $I^{aml}$ be the fixed point of $AM_\mathcal{A}$ and $I_j^{loc}$ the projection of $I^{aml}$ on $DB_j$, $j = 1, ..., n$. Then $I_j^{loc}$ is the fixed point of $R_{DB_j}$.

**Proof Sketch:** It follows from (Subrahmanian, 1994), since AmAU-Datalog programs can be seen as positive GAPs, where update atoms are never evaluated.  □

Intuitively, $I_j$ is the r-interpretation obtained by extracting information about $DB_j$ from $I^{aml}$. The theorem says that every fixed point of the amalgam is an expansion of a corresponding fixed point of a local database, but it does not say that every fixed point of the operator associated with a local database can be expanded in a corresponding fixed point of the amalgam. This result is valid if the supervisor is "strong" (see Definition 9).

When the supervisor is strong, every fixed point of a local database can be expanded to a fixed point of the amalgam, as stated by the following theorem.

*Theorem 4*
Let $DB_1, \ldots, DB_n$ be $n$ AAU-Datalog programs, $S$ a strong supervisor, $\mathcal{A}$ the amalgam of $(S, DB_1, \ldots, DB_n)$, and $I_j^{loc}$ be an r-interpretation. If $I_j^{loc}$ is the fixed point of $R_{DB_j}$, an A-interpretation $I^{aml}$ in the *locale* of $I_j^{loc}$ exists such that $I^{aml}$ is the fixed point of $AM_\mathcal{A}$.

**Proof Sketch:** It follows from (Subrahmanian, 1994), since AmAU-Datalog programs can be seen as positive GAPs, where update atoms are never evaluated.  □

The deductive semantics of a simple transaction $T$ with respect to the amalgam $\mathcal{A} = \cup_{i=1}^n AT(DB_i) \cup C \cup S$ is defined by using the fixpoint operator $AM_P$. As usual in database systems, we give a default set-oriented semantics, that is, the query-answering process computes a set of answers. Before formally introducing the semantics, we need two auxiliary definitions.

*Definition 23*
Given a set of bindings $b$, a transaction $T$, and a substitution $\theta = \{V_1 \leftarrow t_1, \ldots, V_n \leftarrow t_n\}$, we define

- $b_{|T} = \{(X = t) \in b \mid X \text{ occurs in } T\}$;
- $eqn(\theta) = \{V_1 = t_1, \ldots, V_n = t_n\}$. □

In the following, we denote with $\mathsf{Set}(T, \mathcal{A})$ the set of pairs $\langle bindings, updates \rangle$ computed as answers to the simple transaction $T$ (as we will see in Section 5, in a complex transaction, the answers are computed for each transaction in the sequence, therefore the definition has not to be changed).

*Definition 24* (*Query answers*)
Let

- $\mathcal{A} = \cup_{i=1}^{n} AT(DB_i) \cup S \cup C$ be the amalgam of $n$ acceptable AAU-Datalog programs over a complete lattice $\mathcal{T}$, with a strong supervisor $S = SIDB \cup SAR$;
- $IDB_{\mathcal{A}} = \cup_{i=1}^{n}(AT(IDB_{DB_i})) \cup SIDB \cup C$;
- $T = B_1 : [C_1, \mu_1], \ldots, B_n : [C_n, \mu_n] | \alpha_{n+1} D_{n+1} : [E_{n+1}, \mu_{n+1}], \ldots, \alpha_{n+k} D_{n+k} : [E_{n+k}, \mu_{n+k}]$, a simple transaction such that $C_i$ is a NAME-term over $\{s, 1, \ldots, n\}$, $i = 1, \ldots, n$, $E_j$ is a ground NAME-term containing only one element over $\{1, \ldots, n\}$, $\alpha_j \in \{+, -\}$, $j = n+1, \ldots, n+k$.

We define the operator $\mathsf{Set}$ as follows:

$$
\begin{aligned}
\mathsf{Set}(T, P) = \ & \{\langle b, \overline{U} \rangle \mid \\
& A_i : [C_i, \overline{\mu}_i] \leftarrow \alpha_1^i D_1^i : [C_1^i, \mu_1^i], \ldots, \alpha_{k_i}^i D_{k_i}^i : [C_{k_i}^i, \mu_{k_i}^i] \in \mathcal{F}(\mathcal{A}), \\
& \hspace{20em} i = 1, \ldots, n, \\
& \theta = mgu((B_1 : [C_1, \mu_1], \ldots, B_n : [C_n, \mu_n]), (A_1 : [C_1, \overline{\mu}_1], \ldots, \\
& \hspace{20em} A_n : [C_n, \overline{\mu}_n])) \\
& b = eqn(\theta)_{|T} \\
& \overline{U} = \{(\alpha_{n+1} D_{n+1} : [E_{n+1}, \mu_{n+1}])\theta, \ldots, (\alpha_{n+k} D_{n+k} : [E_{n+k}, \\
& \mu_{n+k}])\theta\} \stackrel{+}{\cup} \left( \stackrel{+}{\underset{\substack{i=1,\ldots,n \\ u=1\ldots,k_i}}{\cup}} \{\alpha_u^i D_u^i : [C_u^i, \mu_u^i]\}\theta \right) \}
\end{aligned}
$$

where $\stackrel{+}{\cup}$ is defined in Definition 18. □

*Example 11*
Consider the amalgam presented in Example 6. Figure 4 presents the computation of the fixed point for the considered AmAU-Datalog program. Let $T =? \ pblock(X) : [s, t], tblock(Y, Z) : [s, t]$ be a simple transaction checking if there is a partial block due to any substance and a total block due to any pair of substances between s1,s2 and s3. It is easy to verify that

$$
\begin{aligned}
\mathsf{Set}(T, \mathcal{A}) = \ & \{\langle \{X \leftarrow s3, Y \leftarrow s1, Z \leftarrow s3\}\{+over\_lv(s3) : [1, t], +over\_lv(s3) : [2, t], \\
& +over\_lv(s3) : [3, t], +critical\_lv(s3) : [1, t], +critical\_lv(s3) : [2, t], \\
& +critical\_lv(s3) : [3, t], +partial\_block(s3) : [1, t], +partial\_block(s3) : [2, t], \\
& +partial\_block(s3) : [3, t], +total\_block(s1, s2) : [1, t], \\
& +total\_block(s1, s2) : [2, t], +total\_block(s1, s2) : [3, t].\}\rangle\}
\end{aligned}
$$

◇

$AM_{\mathcal{A}}^1(\emptyset) = \{\ sensor(s1) : [1,t], sensor(s2) : [1,\bot], sensor(s3) : [1,t], over\_lv(s1)[1,t],$
$\qquad\qquad\qquad critical\_lv(s1) : [1,t], over\_lv(s3)[1,t], partial\_block(s3)[1,t], sensor(s1) : [2,\bot],$
$\qquad\qquad\qquad sensor(s2) : [2,t], sensor(s3) : [2,t], over\_lv(s2)[2,t], critical\_lv(s2) : [2,t],$
$\qquad\qquad\qquad over\_lv(s3)[2,t], partial\_block(s3)[2,t], sensor(s1) : [3,t], sensor(s2) : [3,t],$
$\qquad\qquad\qquad sensor(s3) : [3,t], over\_lv(s1)[3,t], over\_lv(s2)[3,t], over\_lv(s3)[3,t],$
$\qquad\qquad\qquad partial\_block(s3)[3,t].\ \}$

$AM_{\mathcal{A}}^2(\emptyset) = AM_{\mathcal{A}}^1(\emptyset) \cup \{\ danger\_lv(s1) : [1,t] \leftarrow +over\_lv(s1) : [1,t], +critical\_lv(s1) : [1,t],$
$\qquad\qquad\qquad danger\_lv(s2) : [1,\bot], danger\_lv(s3) : [1,t] \leftarrow +over\_lv(s3) : [1,t],$
$\qquad\qquad\qquad\qquad\qquad\qquad\qquad\qquad\qquad +critical\_lv(s3) : [1,t],$
$\qquad\qquad\qquad danger\_lv(s1) : [2,\bot], danger\_lv(s2) : [2,t] \leftarrow +over\_lv(s2) : [2,t],$
$\qquad\qquad\qquad\qquad\qquad\qquad\qquad\qquad\qquad +critical\_lv(s2) : [2,t],$
$\qquad\qquad\qquad danger\_lv(s3) : [2,t] \leftarrow +over\_lv(s3) : [2,t], +critical\_lv(s3) : [2,t],$
$\qquad\qquad\qquad danger\_lv(s1) : [3,t] \leftarrow +over\_lv(s1) : [3,t], +critical\_lv(s1) : [3,t],$
$\qquad\qquad\qquad danger\_lv(s2) : [3,t] \leftarrow +over\_lv(s2) : [3,t], +critical\_lv(s2) : [3,t],$
$\qquad\qquad\qquad danger\_lv(s3) : [3,t] \leftarrow +over\_lv(s3) : [3,t], +critical\_lv(s3) : [3,t],$
$\qquad\qquad\qquad danger\_lv(s1) : [\{1,2\},t] \leftarrow +over\_lv(s1) : [1,t], +critical\_lv(s1) : [1,t],$
$\qquad\qquad\qquad danger\_lv(s1) : [\{1,3\},t] \leftarrow +over\_lv(s1) : [1,t], +critical\_lv(s1) : [1,t],$
$\qquad\qquad\qquad\qquad\qquad\qquad\qquad +over\_lv(s1) : [3,t], +critical\_lv(s1) : [3,t],$
$\qquad\qquad\qquad danger\_lv(s1) : [\{2,3\},t] \leftarrow +over\_lv(s1) : [3,t], +critical\_lv(s1) : [3,t],$
$\qquad\qquad\qquad danger\_lv(s2) : [\{1,2\},t] \leftarrow +over\_lv(s2) : [2,t], +critical\_lv(s2) : [2,t],$
$\qquad\qquad\qquad danger\_lv(s2) : [\{1,3\},t] \leftarrow +over\_lv(s2) : [3,t], +critical\_lv(s2) : [3,t],$
$\qquad\qquad\qquad danger\_lv(s2) : [\{2,3\},t] \leftarrow +over\_lv(s2) : [2,t], +critical\_lv(s2) : [2,t],$
$\qquad\qquad\qquad\qquad\qquad\qquad\qquad +over\_lv(s2) : [3,t], +critical\_lv(s2) : [3,t],$
$\qquad\qquad\qquad danger\_lv(s3) : [\{1,2\},t] \leftarrow +over\_lv(s3) : [1,t], +critical\_lv(s3) : [1,t],$
$\qquad\qquad\qquad\qquad\qquad\qquad\qquad +over\_lv(s3) : [2,t], +critical\_lv(s3) : [2,t],$
$\qquad\qquad\qquad danger\_lv(s3) : [\{1,3\},t] \leftarrow +over\_lv(s3) : [1,t], +critical\_lv(s3) : [1,t],$
$\qquad\qquad\qquad\qquad\qquad\qquad\qquad +over\_lv(s3) : [3,t], +critical\_lv(s3) : [3,t],$
$\qquad\qquad\qquad danger\_lv(s3) : [\{2,3\},t] \leftarrow +over\_lv(s3) : [2,t], +critical\_lv(s3) : [2,t],$
$\qquad\qquad\qquad\qquad\qquad\qquad\qquad +over\_lv(s3) : [3,t], +critical\_lv(s3) : [3,t],$
$\qquad\qquad\qquad danger\_lv(s1) : [\{1,2,3\},t] \leftarrow +over\_lv(s1) : [1,t], +critical\_lv(s1) : [1,t],$
$\qquad\qquad\qquad\qquad\qquad\qquad\qquad\qquad +over\_lv(s1) : [3,t], +critical\_lv(s1) : [3,t],$
$\qquad\qquad\qquad danger\_lv(s2) : [\{1,2,3\},t] \leftarrow +over\_lv(s2) : [2,t], +critical\_lv(s2) : [2,t],$
$\qquad\qquad\qquad\qquad\qquad\qquad\qquad\qquad +over\_lv(s2) : [3,t], +critical\_lv(s2) : [3,t],$
$\qquad\qquad\qquad danger\_lv(s3) : [\{1,2,3\},t] \leftarrow +over\_lv(s3) : [1,t], +critical\_lv(s3) : [1,t],$
$\qquad\qquad\qquad\qquad\qquad\qquad\qquad\qquad +over\_lv(s3) : [2,t], +critical\_lv(s3) : [2,t],$
$\qquad\qquad\qquad\qquad\qquad\qquad\qquad\qquad +over\_lv(s3) : [3,t], +critical\_lv(s3) : [3,t],$
$\qquad\qquad\qquad tblock(s1,s2) : [s,t] \leftarrow +total\_block(s1,s2) : [1,t], +total\_block(s1,s2) : [2,t],$
$\qquad\qquad\qquad\qquad\qquad\qquad +total\_block(s1,s2) : [3,t],$
$\qquad\qquad\qquad \ldots\quad \ldots\quad \ldots \qquad\qquad \}$

$AM_{\mathcal{A}}^3(\emptyset) = AM_{\mathcal{A}}^2(\emptyset) \cup \{\ pblock(s3) : [s,t] \leftarrow +over\_lv(s3) : [1,t], +over\_lv(s3) : [2,t],$
$\qquad\qquad\qquad\qquad\qquad\qquad +over\_lv(s3) : [3,t], +critical\_lv(s3) : [1,t],$
$\qquad\qquad\qquad\qquad\qquad\qquad +critical\_lv(s3) : [2,t], +critical\_lv(s3) : [3,t],$
$\qquad\qquad\qquad\qquad\qquad\qquad +partial\_block(s3) : [1,t], +partial\_block(s3) : [2,t],$
$\qquad\qquad\qquad\qquad\qquad\qquad +partial\_block(s3) : [3,t].\ \}$

$AM_{\mathcal{A}}^4(\emptyset) = AM_{\mathcal{A}}^3(\emptyset) = \mathcal{F}(\mathcal{A})$

Figure 4. The fixpoint computation

### 4 Active part semantics

After computing the deductive semantics of amalgam, active rules have to be considered in order to determine which additional updates they generate. The active part semantics is directly defined for the amalgam and is given following the line of the PARK semantics proposed in (Gottlob *et al.*, 1996).

In order to introduce such semantics, three main aspects have to be considered:

- First of all, we have to determine which active rules are fired. An active rule is fired if all update atoms contained in its body have already been generated either by a deductive rule or by another active rule and all the deductive atoms can be deduced from the deductive rules.
- After deciding which rules are fired, a mechanism has to be proposed to deal with conflicts, i.e., insertion and deletion of the same information. In particular, conflicts have to be detected and removed by using a conflict resolution function.
- Finally, we have to specify how the overall triggering process takes place, since updates generated by an active rule can trigger a different active rule and this process can be iterated. Conditions to guarantee termination have also to be presented.

In the following, all the previous aspects are discussed in details.

#### *4.1 Validity conditions*

An active rule is fired if the following conditions hold:

- each update atom in the event part of the active rule has already been generated either by the deductive part or by firing another active rule;
- each positive atom in the condition part of the active rule is satisfied by the deductive part semantics;
- each negative atom in the condition part of the active rule is not satisfied by the deductive part semantics.

The previous conditions point out that validity of event-condition atoms has to be checked with respect to something different than r-interpretations, since also update atoms have to be considered. To this purpose, the concept of *restricted-intermediate interpretation* (ri-interpretation) is provided.

*Definition 25* (*ri-interpretation*)
Let $\mathcal{B}^{\pm}$ be the set of ground, non annotated atoms that can be constructed over a base language $\mathcal{L}$. Let $\mathcal{T}$ be a complete lattice. An ri-interpretation is a function associating with each element of $\mathcal{B}^{\pm}$ a principal of $\mathcal{T}$. We denote with $\mathbf{RI}(\mathcal{T})$ the set of all ri-interpretations over $\mathcal{T}$. □

Similarly to r-interpretation, also ri-interpretations can be seen as a set of amalgamated (but not constrained) atoms. In the following, for the sake of simplicity, we use this set-based notation.

30  *E. Bertino, B. Catania, and P. Perlasca*

To establish when an active rule can be triggered, we introduce a *validity* predicate. Given an amalgamated atom and an ri-interpretation, the validity predicate returns true if the atom is valid in the considered ri-interpretation.

*Definition 26* (*Validity*)
Let $a$ be a ground amalgamated literal. $a$ is valid in an ri-interpretation $I$ (denoted by $valid(a, I)$) if one of the following conditions holds:

| | |
|---|---|
| $(I \cap \{A : [D, \overline{\mu}], +A : [D, \overline{\mu}]\}) \neq \emptyset, \mu \leq \overline{\mu}$ | and $a = A : [D, \mu]$ |
| $(I \cap \{A : [D, \overline{\mu}], +A : [D, \overline{\mu}]\}) = \emptyset$ or $-A : [D, \overline{\mu}] \in I, \mu \leq \overline{\mu}$ | and $a = \neg A : [D, \mu]$ |
| $+A : [D, \overline{\mu}] \in I, \mu \leq \overline{\mu}$ | and $a = +A : [D, \mu]$ |
| $-A : [D, \overline{\mu}] \in I, \mu \leq \overline{\mu}$ | and $a = -A : [D, \mu]$ |

□

According to the previous definition, a positive amalgamated $(\Pi^e \cup \Pi^i, \Sigma, V)$-atom is valid in $I$ if $I$ contains the same atom with a greater annotation or if an atom with a greater annotation has to be inserted. A negative amalgamated $(\Pi^e \cup \Pi^i, \Sigma, V)$-atom is valid in $I$ if $I$ contains an update deleting the same atom with a greater annotation or if the corresponding positive atom is not valid. An amalgamated $(\Pi^u, \Sigma, V)$-atom is valid in $I$ if $I$ contains the same atom with a greater annotation. Notice that both $A : [D, \mu]$ and $\neg A : [D, \mu]$ can be valid according to this definition. The intuition behind the above definition is that since a positive or negative atom belongs to the condition part of the active rule, its validity must be checked with respect to the derived atoms and also to the inserted and deleted atoms. Otherwise, to represent the occurrence of an event, we require that just the update modeling such an event belongs to the ri-interpretation.

An active rule is therefore triggered by an ri-interpretation $I$ if all its event-condition atoms are true with respect to $I$.

*Definition 27*
Let $I$ be an ri-interpretation and $r$ an AmAU-Datalog active rule. $r$ is triggered by $I$ if all the event-condition update atoms and literals in $r$ are valid in $I$.   □

*Example 12*
Let $I$ be an ri-interpretation such that $+p_2(a, b) : [1, 0.75] \in I, p_1(a, b) : [1, 0.75] \in I$. Let $r, s$ be the following AmAU-Datalog active rules:
$r : \quad +p_2(a, b) : [1, 0.75] |\ p_1(a, b) : [1, 0.5] \rightarrow +p_1(a, b) : [1, 0.75]$;
$s : \quad +p_2(a, b) : [1, 1] |\ p_1(a, b) : [1, 0.5] \rightarrow -p_1(a, b) : [1, 1]$.
$r$ is triggered by $I$ since all the event-condition atoms in $r$ are valid in $I$. More precisely, $p_1(a, b) : [1, 0.5]$ is valid since $p_1(a, b) : [1, 0.75] \in I$ and $0.5 < 0.75$ whereas $+p_2(a, b) : [1, 0.75]$ is valid since it belongs to $I$. On the other hand, $s$ is not triggered by $I$ since $1 > 0.75$.   ◇

### 4.2 Blocked rule instances

Suppose that, given an ri-interpretation, more than one rule is fireable. It could happen that the actions (i.e., the updates) to be executed are conflicting. This happens when some active rules add a certain atom and some others remove it. The concept of conflicting updates can be formally defined as follows.

*Definition 28* (*Annotated conflicting update atoms*)
Let $U_1 = \alpha_1 D : [i, \mu_1]$ and $U_2 = \alpha_2 D' : [i, \mu_2]$ be two annotated update atoms. $U_1$ and $U_2$ are conflicting if there exists a substitution $\sigma$ such that $(D)\sigma = (D')\sigma$ and $\alpha_i \in \{+, -\}$, $i = 1, 2$, $\alpha_1 \neq \alpha_2$. □

Note that two update atoms are conflicting if they require the insertion and the deletion of the same information, independently from the associated annotations.

An ri-interpretation is *consistent* if it does not contain any pair of *conflicting updates*. For what we will do in the following, it is important not only to identify conflicting updates but also the rule instances (also called *rule grounding*) generating them. To this purpose, we introduce the concept of *conflict*.

*Definition 29* (*Conflicts*)
A pair $(r, \theta)$, where $r$ is a rule and $\theta$ is a ground substitution for $r$ is called a *rule grounding*.

Let $P$ be a set of AmAU-Datalog active rules and $I$ an ri-interpretation for $P$. Then conflicts$(P, I)$ is a set of maximal tuples of the form $(i, A, \mathsf{ins}, \mathsf{del})$ such that $i$ is a database identifier, $A$ is a ground atom, and ins and del are sets of active rule groundings. For each such triple the following conditions must hold:

1. there exists r,r'∈P

$$r = A_1 : [D_1, \mu_1], \ldots, A_n[D_n, \mu_n] \to B_1 : [E_1, \psi_1], \ldots, B_m : [E_m, \psi_m]$$

$$r' = A'_1 : [D'_1, \mu'_1], \ldots, A'_{n'}[D'_{n'}, \mu'_{n'}] \to B'_1 : [E'_1, \psi'_1], \ldots, B'_{m'} : [E'_{m'}, \psi'_{m'}]$$

where $A_h, h = 1, ..., n$, $A'_{h'}, h' = 1, ..., n'$ are either amalgamated update atoms or deductive amalgamated literals, $B_k, k = 1, ..., m$, $B_{k'}, k' = 1, ..., m'$ are update atoms, and $\theta, \theta'$ ground substitutions such that
   - $\forall h\ 1 \leq h \leq n$, $\mathsf{valid}(A_h : [D_h, \mu_h]\theta, I)$,
   - $\forall k\ 1 \leq k \leq n'$, $\mathsf{valid}(A'_k : [D'_k, \mu'_k]\theta', I)$,
   - $\exists p, q.\ 1 \leq p \leq m.1 \leq q \leq m'.\ B_p : [E_p, \psi_p] = +C : [E_p, \psi_p], B'_q : [E'_q, \psi'_q] = -F : [D'_q, \psi'_q]$ and $A = C\theta = F\theta'$.

2. For all $r$, $r'$ and $\theta$, $\theta'$, satisfying condition 1 above, $(r, \theta) \in \mathsf{ins}$ and $(r', \theta') \in \mathsf{del}$.

A tuple $(i, A, \mathsf{ins}, \mathsf{del}) \in \mathsf{conflicts}(P, I)$ is called a *conflict*. □

To solve conflicts, a parametric conflict resolution policy is introduced. Such a policy specifies, for each conflict, which update must prevail.

*Definition 30* (*Conflict resolution policy*)
Given an AmAU-Datalog extensional database *EDB*, a set of rules $P$, an ri-interpretation $I$ and a conflict $c$, we define $\mathsf{sel}(EDB, P, I, c)$ as a total function with range $\{\mathsf{insert}, \mathsf{delete}\}$. □

The intended meaning of $\mathsf{sel}(EDB, P, I, (i, A, \mathsf{ins}, \mathsf{del}))$ is to choose whether atom $A$, object of the conflict, should be inserted in or deleted from $I$, thus effectively choosing which of the conflicting update requests should prevail. Note that the selection function cannot require the insertion and the deletion of the same atom, since for each ground atom only one conflict can exist.

Gottlob et al. (Gottlob *et al.*, 1996) present a number of commonly adopted policies, and discuss their advantages and disadvantages. We briefly recall here some of them. The *principle of inertia* states that *both* the conflicting updates should be discarded, thus leaving EDB in the same state as before with respect to $db\!:\!a$ (in our framework, this can be obtained by returning insert if $db\!:\!a$ was already in EDB, delete otherwise). The *source priority* policy determines which update should prevail according to which database the rules requesting such updates come from (in our framework, this can be obtained by using the mapping $f$ which establishes the relation between rules and databases of our system). The *rule priority* policy, found in systems such as Ariel (Hanson, 1996), Postgres (Stonebraker *et al.*, 1990) and Starburst (Widom and Finkelstein, 1990), assumes that each rule has a (static or dynamic) priority associated with it; sel returns insert or delete as needed to preserve the update requested by the highest-priority rule. Other policies, like voting schemes or user queries, are also reasonable, but the final choice is left to the particular application.

Based on the result of the sel policy, we prevent the rule instances in one of the two sets of a conflict from firing, by blocking them according to the following definition. Blocked rule instances will then be used in the next subsection to specify how active rule computation takes place.

*Definition 31* (*Blocked rule instances*)

Given an AmAU-Datalog extensional database *EDB*, a set of rules $P$, a conflict resolution policy sel, and an ri-interpretation $I$, let

$$\begin{aligned}
X &= \{\text{del} \mid (i, A, , \text{ins}, \text{del}) \in \text{conflicts}(P, I) \text{ and} \\
&\quad \text{sel}(EDB, P, I, (i, A, \text{ins}, \text{del})) = \text{insert}\} \\
Y &= \{\text{ins} \mid (i, A, \text{ins}, \text{del}) \in \text{conflicts}(P, I) \text{ and} \\
&\quad \text{sel}(EDB, P, I, (i, A, \text{ins}, \text{del})) = \text{delete}\}.
\end{aligned}$$

We define

$$\text{blocked}(EDB, P, I, \text{sel}) = \left(\bigcup_{x \in X} x\right) \cup \left(\bigcup_{y \in Y} y\right). \qquad \Box$$

We block an entire rule instance, rather than a single update, so that the set of updates requested by the same rule instance exhibits an atomic behavior: either all the updates in the set are executed, or no update at all. This avoids the risk of making the database inconsistent due to partially-executed actions.

*Example 13*

Consider the AmAU-Datalog active rules presented in Example 12 and let $I$ be an ri-interpretation such that: $+p_2(a, b) : [1, 1] \in I, p_1(a, b) : [1, 0.75] \in I$.

In this situation, both $r$ and $s$ are triggered by $I$ since all their event-condition atoms are valid in $I$ and a conflict $C = (1, p_1(a, b), \{(r, \emptyset)\}, \{(s, \emptyset)\})$ arises. If we assume that the conflict resolution function privileges deletions, we block the rule instance $r$, obtaining $blocked(EDB, P, I, \text{sel}) = \{r\}$. $\diamond$

*4.3 Computation*

Using the above concepts, given a set of AmAU-Datalog rules $P$, a set of blocked rule instances $B$, and an ri-interpretation $I$, we define an immediate consequence operator over ri-interpretations $\Gamma_{P,B}(I)$, similarly to other bottom-up operators defined in the logic programming context. However, differently from them, some rules may not be fired during the computation, even if their body is valid, due to the blocked set of rules.

In performing such a computation, we have to consider not only active rules but also deductive rules, since we have to check validity of both the event and the condition part of active rules. In particular, the following aspects should be taken into account:

- conditions should be checked by taking into account the requested updates;
- the resolution of a condition should not affect the state of the system.

While the first condition is assured by the considered notion of validity (see Definition 26), to fulfill the second condition we remove the update part from the rules of the intensional databases by using the purification operation defined below.

*Definition 32* (*Purification*)
Given the intensional database IDB of an AmAU-Datalog program, we define its *purified* version $\widehat{\text{IDB}}$ as the set of rules

$$B_1, \ldots, B_m \to H.$$

such that there exists in IDB a rule

$$H \leftarrow U_1, \ldots, U_n, B_1, \ldots, B_m. \qquad \Box$$

It is worth noting that a query is provable in $\widehat{IDB_\mathcal{A}} \cup EDB_\mathcal{A}$ if and only if it is provable in $IDB_\mathcal{A} \cup EDB_\mathcal{A}$, with the same computed answers. The purification only avoids the side effects of the query evaluation. Also notice that we reversed the direction of the arrow in order to have a uniform notation with active rules.

In the sequel, we generically use the term "rules" to refer to both active and purified amalgamated rules.

*Definition 33* (*Immediate consequence operator*)
Given a set of AmAU-Datalog rules $P$, a set of blocked rule instances $B$, and an ri-interpretation $I$, we define $\Gamma_{P,B}(I)$ as $\sqcup U$,[9] where $U$ is the smallest set satisfying the following conditions:

1. $I \subseteq U$;
2. If $r = A_1 : [D_1, \mu_1], \ldots, A_n[D_n, : \mu_n] \to B_1 : [E_1, \psi_1], \ldots, B_m : [E_m, \psi_m]$, $r \in P$ and $\theta$ is a ground substitution such that
   - $(r, \theta) \notin B$;
   - $\mathsf{valid}(A_k : [D_k, \mu_k]\theta, I)$, $k = 1, \ldots, n$, then $\{(B_1 : [E_1, \psi_1])\theta, \ldots, (B_m : [E_m, \psi_m])\theta\} \subseteq U$. $\qquad \Box$

---

[9] $\sqcup U = \{A : [D, \mu] | A : [D, \mu_1], \ldots, A : [D, \mu_n] \in U \text{ and } \sqcup_i \mu_i = \mu\}$.

The main difference of the above operator with respect to the traditional immediate consequence operator of logic programming is that it may happen that some of the rules are not fired even if their body is valid, due to the blocked set of rules. Moreover, such operator is monotonic but it is not continuous.[10] Indeed, if we consider an AmAU-Datalog program $P$ without update atoms and active rules and we let $B = \{\}$, then $\Gamma_{P,B}$ coincides with $R_P$. Thus, from Theorem 1, it follows that $\Gamma_{P,B}$ is not continuous. However, the following proposition shows that if $P$ is c-annotated, or the lattice is finite, $\Gamma_{P,B}$ admits a fixpoint, reachable in a finite number of steps.

*Proposition 2*
Let $P$ be an AmAU-Datalog database and $B$ a set of blocked rule instances. If $\mathcal{T}$ is finite or $P$ is c-annotated, operator $\Gamma_{P,B}$ admits a fixpoint $lfp(\Gamma_{P,B}) = \Gamma_{P,B} \uparrow \omega$ and there exists $k$ such that $\Gamma_{P,B} \uparrow \omega = \Gamma_{P,B} \uparrow k$.

*Proof*
If $\mathcal{T}$ is finite, the number of ri-interpretations is finite. Therefore $\Gamma_{P,B}$ is a monotonic operator over the finite lattice of ri-interpretations, ordered by $\subseteq^+$, where $\subseteq^+$ is the usual containment between sets with the following exception: if $I$ contains $A : [D, \mu]$, then $\{A : [D, \mu']\} \subseteq^+ I$, for all $\mu' \leq \mu$. Thus, it is also continuous and the thesis follows. If $P$ is c-annotated, it is also acceptable and $lfp(\Gamma_{P,B}) = \Gamma_{P,B} \uparrow \omega$ follows from a reasoning similar to that presented in the proof of Lemma 1. Moreover, since annotations are fixed, the number of terms that can be constructed over the base language is finite, and rules are range restricted, thus there exists $k$ such that $\Gamma_{P,B} \uparrow \omega = \Gamma_{P,B} \uparrow k$. □

In general, the application of the function $\Gamma_{P,B}$ to a consistent ri-interpretation does not return a consistent ri-interpretation, as shown in (Bertino *et al.*, 1998). Therefore, even under the conditions presented in Proposition 2, we cannot compute the semantics of $P$ as the least fixpoint of $\Gamma_{P,B}$. We must instead appropriately select rules, that is, we must build a set of blocked rules $B$ such that the least fixpoint of $\Gamma_{P,B}$ is consistent. Thus, instead of dealing with ri-interpretations, the notion of bi-structures is introduced, as in (Gottlob *et al.*, 1996), in order to take blocked rules into account during the computation.

*Definition 34* (*Bi-structures*)
A bi-structure $\langle B, I \rangle$ consists of a set $B$ of rule groundings and of an ri-interpretation $I$. We define an order relation on bi-structures as follows:

$$\langle B, I \rangle \prec \langle B', I' \rangle \quad \stackrel{\text{def}}{\Leftrightarrow} \quad \begin{cases} B \subset B' & \text{or} \\ B = B' \text{ and } I \subset^+ I' \end{cases}$$

Given $\mathcal{A}$ and $\mathcal{B}$ bi-structures, $\mathcal{A} \preceq \mathcal{B} \quad \equiv \quad (\mathcal{A} = \mathcal{B} \vee \mathcal{A} \prec \mathcal{B})$. □

On this domain, we can define an operator having a fixpoint, that is used to compute the semantics of the active part.

*Definition 35* ($\Delta$ *operator*)

---

[10] In Active U-Datalog, a similar operator has been defined which however is continuous (Bertino *et al.*, 1998).

Given a set of AmAU-Datalog rules $P$, a bi-structure $\langle B, I \rangle$ and a conflict resolution policy sel, we define

$$\Delta_{P,\mathsf{sel}}(\langle B, I \rangle) = \begin{cases} \langle B, \Gamma_{P,B}(I) \rangle & \text{if } \Gamma_{P,B}(I) \text{ is consistent;} \\ \langle B \cup \mathsf{blocked}(I^e, P, I, \mathsf{sel}), I^e \rangle & \text{otherwise.} \end{cases}$$

where $I^e$ is the set of the extensional amalgamated atoms contained in $I$, that is: $I^e = \{A : [D, \mu] \in I \mid pred(A) \in \Pi^e\}$. □

The definition of $\Delta$ we give here differs from the original in (Gottlob *et al.*, 1996) because the set of rules $P$ contains not only rules with updates in the right hand side (properly active rules) but also purified rules that allow us to derive intensional knowledge rather than new updates. Notice that this extension does not affect the consistency of ri-interpretations, since the purified rules can only add amalgamated ($\Pi^i, \Sigma, V$)-atoms to the ri-interpretation.

The intuitive idea of the $\Delta$ operator is that, if no conflict arises, $\Delta$ does not change the blocked rule set $B$, and only the ri-interpretation of the bi-structure is changed by adding the immediate consequences of the non blocked rules. On the other hand, as soon as a conflict arises, the conflict is solved via the resolution policy sel and all blocked rule instances are collected. Then, the computation of $\Delta$ is started again from the ri-interpretation $I^e$ with the augmented set of blocked rules. The ri-interpretation $I^e$ represents the set of the extensional atoms of the database, and we have to resort to it to be sure that the starting point of the new computation does not contain atoms whose validity depends on actions of rule instances that are now blocked.

In order to define a semantics based on the $\Delta$ operator, we must be sure that a fixed point of $\Delta$ exists. However, as shown in (Bertino *et al.*, 1998), $\Delta$ in general is not continuous, due to the non-continuity of $\Gamma_{P,B}$ and to function blocked, which in turns depends on an arbitrary function $sel$ and is not monotonic. Therefore, we cannot prove that $\Delta$ has a fixpoint by using the fixpoint theorem. However, if the program is c-annotated or if the lattice is finite, it is possible to prove the following result.

*Proposition 3*
Let $P$ be a set of AmAU-Datalog rules, sel a conflict resolution policy, $\mathcal{D} = \langle B, I \rangle$ a bi-structure, $I$ a set of ground amalgamated extensional atoms. Suppose that either $P$ is c-annotated or the lattice is finite. The following statements hold:

1. $\mathcal{D} \preceq \Delta_{P,\mathsf{sel}}(\mathcal{D})$,
2. $\Delta_{P,\mathsf{sel}}^{\omega}(\mathcal{D})$ is a fixpoint of $\Delta_{P,\mathsf{sel}}$ and there exists $k$ such that $\Delta_{P,\mathsf{sel}}^{\omega}(\mathcal{D}) = \Delta_{P,\mathsf{sel}}^{k}(\mathcal{D})$.

*Proof*
1. Let $\Delta_{P,\mathsf{sel}}(\mathcal{D}) = \langle B', I' \rangle$. If $I'$ is consistent, then $B' = B$ and $I' = \Gamma_{P,B}(I) \supseteq^+ I$ by definition of $\Gamma$; hence $\langle B, I \rangle \preceq \langle B', I' \rangle$. If instead $I'$ is not consistent, then $B' = B \cup \mathsf{blocked}(EDB, P, I, \mathsf{sel}) \supseteq B$ and so we have again $\langle B, I \rangle \preceq \langle B', I' \rangle$.
2. By statement 1, for all natural numbers $n$, we have

$$\Delta_{P,\mathsf{sel}}^{n}(\mathcal{D}) \preceq \Delta_{P,\mathsf{sel}}(\Delta_{P,\mathsf{sel}}^{n}(\mathcal{D})).$$

Now suppose that the lattice is finite. Under this hypothesis, bi-structures are finite and form a complete lattice. Hence, $\{\Delta_{P,\mathsf{sel}}^{i}(\mathcal{D})\}_{i \in \mathbb{N}}$ is a chain in the cpo of the bi-structures. Since such a cpo is finite, every chain consists of a finite number of elements. Therefore

$$\exists k. \forall n \geq k.\ \Delta_{P,\text{sel}}^n(\mathcal{D}) = \Delta_{P,\text{sel}}^{n+1}(\mathcal{D}).$$

We can conclude that $\Delta_{P,\text{sel}}^k(\mathcal{D})$ is a fixpoint of $\Delta_{P,\text{sel}}$. If the lattice is not finite but $P$ is c-annotated, it follows that: (i) the number of possible sets of blocked instances is finite, since annotations are fixed, the number of terms that can be constructed over the base language is finite, and rules are range restricted; (ii) due to Proposition 2, $\Gamma_{P,B}$ admits a fixpoint $lpf(\Gamma_{P,B}) = \Gamma_{P,B} \uparrow \omega = \Gamma_{P,B} \uparrow k_1$. From fact (i), it follows that there exists $k_2$ such that for all $h \geq k_2$ $\Delta_{P,\text{sel}}^{k_2}(\mathcal{D}) = \langle B, I \rangle$, $\Delta_{P,\text{sel}}^h(\mathcal{D}) = \langle B, I' \rangle$ and $I' \subseteq^+ I$. From (ii) and the definition of $\Delta$, we obtain the thesis. □

The next theorem shows that we can find a set of blocked rules $B$ such that the least fixpoint of $\Gamma_{P,B}$ is a consistent ri-interpretation.

*Theorem 5*
Let $P$ be a set of AmAU-Datalog rules, sel a conflict resolution policy, $\mathcal{D} = \langle B, I \rangle$ a bi-structure, $I$ a set of ground amalgamated extensional atoms. Suppose that either $P$ is c-annotated or the lattice is finite. Then, there exists $k$ such that $\Delta_{P,\text{sel}}^k(\mathcal{D})$ is a fixpoint of $\Delta_{P,\text{sel}}$ and if $\Delta_{P,\text{sel}}^k(\mathcal{D}) = \langle B', I' \rangle$, then $I' = lfp_I(\Gamma_{P,B'})$[11] and $I'$ is consistent. Moreover, it contains a minimal set of blocked rule groundings.

*Proof*
First we notice that if $\Delta_{P,\text{sel}}(\langle B_1, I_1 \rangle) = \langle B_2, I_2 \rangle$ then $I_1{}^\bot = I_2{}^\bot$, that is the set of ground extensional atoms is not modified by $\Delta$. This follows from the definition of $\Delta_{P,\text{sel}}$ and from the fact that $\Gamma_{P,B}$ can add only intensional and update atoms to an ri-interpretation. As a consequence, for all natural numbers $n$, if $\Delta_{P,\text{sel}}^n(\langle B_1, I_1 \rangle) = \langle B_{n+1}, I_{n+1} \rangle$ then $I_1{}^\bot = I_{n+1}{}^\bot$.
Since the ri-interpretation of $\mathcal{A}$ only consists of extensional atoms, then $I^\bot = I$.

By Lemma 3, $\Delta_{P,\text{sel}}^k(\mathcal{A})$ is a fixpoint of $\Delta_{P,\text{sel}}$. By definition of $\Delta$ and by the above remark, there exists $i \leq k$ such that $\Delta_{P,\text{sel}}^i(\mathcal{D}) = \langle B', I \rangle$. Then for all $j$ such that $i \leq j \leq k$, we have $\Delta_{P,\text{sel}}^j(\mathcal{D}) = \langle B', \Gamma_{P,B'}^{j-i}(I) \rangle$, because $B'$ does not increase. Since $\Delta_{P,\text{sel}}^k(\mathcal{D}) = \langle B', I' \rangle = \langle B', \Gamma_{P,B'}^{k-i}(I) \rangle$ is a fixpoint, then $\langle B', \Gamma_{P,B'}^{k-i}(I) \rangle = \langle B', \Gamma_{P,B'}^{k-i+1}(I) \rangle$. Therefore $I' = \text{lfp}_I(\Gamma_{P,B'})$ and by definition of $\Delta$, the set $I'$ is consistent (otherwise the set of blocked rules would be augmented).

In order to show that the computed set of blocked rules is minimal, suppose, by contradiction, that the computed set of blocked rule instances - say $B$ - is not minimal. This means that there exists a set of rule groundings $R_1 = \{r_1, ..., r_n\} \in B$ and a set of rule groundings $R_2 = \{r'_1, ..., r'_m\}$, $m < n$, $R_1 \cap R_2 = \emptyset$, such that $lfp_I(\Gamma_{P,(B-R_1) \cup R_2})$ is consistent. Two cases may arise:

- Suppose that $R_2 \subseteq B$. This means that rule groundings in $R_1$ does not lead to any conflict. But this not possible, since they have been selected by using the selection function.

---
[11] $lfp_I(f)$ denotes the least fixpoint of $f$ which is greater than or equal to $I$.

- Suppose that $R_2 \not\subseteq B$. This means that by blocking rule groundings in $R_1$, rule groundings in $R_2$ cannot be generated. Now suppose not to block rule groundings in $R_1$. These rules have been selected by the selection function and therefore they generate conflicts. Moreover, their generation does not depend on the generation of rule groundings in $R_2$, therefore by blocking rule groundings in $R_2$, rule groundings in $R_1$ are not blocked. This also means that $lfp_I(\Gamma_{P,(B-R_1)\cup R_2})$ cannot be consistent.

In both cases, we arrive to a contradiction, therefore the computed set of blocked rule groundings in minimal. □

## 5 Integrating deductive and active semantics

In this section we show how the deductive and active semantics presented above fit together and how the result of a transaction is computed.

We are interested in modeling as observable property of a transaction the following information: the set of answers, the database state, and the result of the transaction itself (that is, Commit or Abort).

*Definition 36* (*Observables*)
An *observable* is a triple $\langle \mathsf{Ans}, \mathit{EDB}, \mathsf{Res} \rangle$ where Ans is a set of bindings, *EDB* is an extensional database and Res $\in$ {Commit, Abort}. The set of observables is Oss. □

The semantics presented in Sections 3 and 4 does not include the execution of the collected updates, neither considers the transactional behavior. We now define a function which, given an i-interpretation and the current state of the system, returns the next state obtained by executing the updates in the i-interpretation.

*Definition 37* (*Updates incorporation*)
Given an ri-interpretation $I$ and a AmAU-Datalog extensional database $EDB = EDB_1 \cup \ldots \cup EDB_n$, we define

$$\mathsf{incorp}(I, EDB) = \bigcup_i (EDB_i \setminus^{inc} \{A : [i, \overline{\mu}] \mid -A : [i, \overline{\mu}] \in I\}) \bigcup{}^+ \{A : [i, \overline{\mu}] \mid +A : [i, \overline{\mu}] \in I\}.$$

where $A \setminus^{inc} B$ is defined as in Definition 18. $A \setminus^{inc} B$ is defined as the usual set difference with the following exception: if $A$ contains $D : [i, \mu_1]$ and $B$ contains $D : [i, \mu_2]$, then:

- if $\mu_1 < \mu_2$, $A \setminus^{inc} B$ does not contain $D : [i, \mu_1]$;
- if $\mu_1 > \mu_2$, $A \setminus^{inc} B$ contains $D : [i, \mu_1]$. □

Informally, update incorporation is based on the following rules. Suppose that an annotated atom $A_1 : [i, \mu_1]$ has to be inserted/deleted and $EDB_i$ already contains an annotated atom $A_1 : [i, \mu_2]$. The following cases may arise:

- $\mu_1 = \mu_2$: the insertion of an already present fact is required, then the extensional database does not change;
- $\mu_1 > \mu_2$: since the truth value of $\mu_1$ is stronger, the fact $A : [i, \mu_1]$ prevails and the new extensional database contains $A : [i, \mu_1]$ in place of $A : [i, \mu_2]$;

- $\mu_1 < \mu_2$: $A : [i, \mu_2]$ is maintained since the truth value $\mu_2$ already includes the truth value $\mu_1$, in other words the insertion of an already present fact is required;
- $\mu_1$ and $\mu_2$ non comparable: $A : [i, \mu_2]$ is replaced by $A : [i, \sqcup\{\mu_1, \mu_2\}]$.

Now consider deletion:

- $\mu_1 = \mu_2$: the new extensional database does not contain the fact $A : [i, \mu_2]$;
- $\mu_1 > \mu_2$: since the truth value of $\mu_1$ is stronger, the request of deletion prevails and the new extensional database does not contain fact $A : [i, \mu_2]$;
- $\mu_1 < \mu_2$: $A : [i, \mu_2]$ is maintained since the truth value $\mu_1$ is not sufficiently strong to delete such fact;
- $\mu_1$ and $\mu_2$ non comparable: since it is not possible to establish which of the two values is the strongest, the extensional database is left unchanged.

The semantics of an AmAU-Datalog program can now be defined as follows.

*Definition 38* (*AmAUDatalog semantics*)
Given an AmAU-Datalog program P, $P = (S, DB_1, ..., DB_n)$, a transaction $T$ and a conflict resolution policy sel, if $P$ is c-annotated or the lattice is finite, the semantics of a transaction $T$ in $P$ is denoted by the function *Sem* defined as

$$Sem_{P,\text{sel}}(T) = \mathcal{S}_{\text{IDB,AR,sel}}(T)(\langle \emptyset, EDB, \text{Commit}\rangle)$$

where the function $\mathcal{S}_{\text{IDB,AR,sel}}(T) : \text{Oss} \to \text{Oss}$ is defined as follows:
If $T$ is a simple transaction, then

$$\mathcal{S}_{\text{IDB,AR,sel}}(T)(\langle \alpha, \varepsilon, \rho\rangle) = \begin{cases} \langle \emptyset, \varepsilon, \text{Abort}\rangle & \text{if } \rho = \text{Abort} \\ \langle \text{Ans}, \text{incorp}(I, \varepsilon), \text{Commit}\rangle & \text{if } \rho \neq \text{Abort} \text{ and } \overline{U} \text{ is ground} \\ \langle \emptyset, \varepsilon, \text{Commit}\rangle & \text{otherwise} \end{cases}$$

where

$$\begin{aligned}
\text{Ans} &= \alpha \circ \{b_j \mid \langle b_j, \overline{u}_j\rangle \in \text{Set}(T, P)\}, \text{ whereas } \circ \text{ is the concatenation operator} \\
\overline{U} &= \bigcup \{\overline{u}_j \mid \langle b_j, \overline{u}_j\rangle \in \text{Set}(T, P)\} \\
\langle B, I\rangle &= \Delta^\omega_{\Xi,\text{sel}}(\langle \emptyset, \varepsilon\rangle) \\
\Xi &= AR \cup \widehat{IDB_P} \cup \{\to +A : [i, \mu] \mid +A : [i, \mu] \in \overline{U}\} \cup \\
&\quad \{\to -A : [i, \mu] \mid -A : [i, \mu] \in \overline{U}\}.
\end{aligned}$$

If $T$ is a complex transaction $T_1; \ldots ; T_k$ ($k \geq 2$), then

$$\mathcal{S}_{\text{IDB,AR,sel}}(T_1; \ldots ; T_k)(Oss) = \mathcal{S}_{\text{IDB,AR,sel}}(T_2; \ldots ; T_k)(\mathcal{S}_{\text{IDB,AR,sel}}(T_1)(Oss))$$

where $T_1, \ldots, T_k$ are simple transactions. □

To compute the semantics of a simple transaction, first we build the set of answers in the marking phase (Definition 24). This step returns a set of bindings for the variables (both in $V$ and in $\mathcal{V}$) of the transaction (Ans) and a set of amalgamated updates ($\overline{U}$) which are requested but which will not necessarily be executed. Then, we gather rules in order to apply the $\Delta$ operator (Definition 35).

Such a set of rules ($\Xi$) contains the rules in *AR* and the amalgamated updates requested from the deductive part ($\overline{U}$), represented as rules with neither event nor condition. The updates in $\overline{U}$ become the initial events to which the active rules in *AR* have to react.

To obtain the set of updates to be executed, we apply the $\Delta$ operator starting from an empty set of blocked rule instances and from the extensional database as initial ri-interpretation. Theorem 5 assures that a fixpoint of $\Delta_{\Xi,\text{sel}}$ is reached in a finite number of steps by computing the approximations $\Delta^i_{\Xi,\text{sel}}(\langle \emptyset, \varepsilon \rangle)$ and that the ri-interpretation, $I$, in the resulting bi-structure is consistent. Finally, the new state of the database is computed by incorporating the updates belonging to $I$ in the current state of the database, following Definition 37.

The semantics of a complex transaction is simply given by the sequential composition of the semantics of its components. The state of the system is updated after each simple transaction. Besides, we return a list of sets of answers, one for each transaction composing the complex one.[12]

It is important to note that the proposed semantics generates abort only due to media failures. As already noted, the answer set Set and the $\Delta$ fixpoint are computed in a finite number of steps, hence $Sem_{P,\text{sel}}$ is computed in a finite number of steps.

*Example 14*

Consider Example 3. In order to show how the semantics of AmAU-Datalog programs works, consider the transaction $T =? \ pblock(X) : [s,t], tblock(Y,Z)[s,t]$, executed against the amalgam presented in Example 6, whose fixpoint has been presented in Figure 4. In order to make clearer the computation, we use a set-based representation of ri-interpretations. In order to compute $Sem_{P,sel_\mathcal{A}}(T)$, suppose that our selection function privileges deletions. Now, let $\Xi = \widehat{IDB_\mathcal{A}} \cup AR_\mathcal{A} \cup USet(T,\mathcal{A})$, where $USet(T,\mathcal{A})$ represents the set of rules generated from the updates contained in $\text{Set}(T,\mathcal{A})$, presented in Example 11, i.e.

$USet(T,\mathcal{A}) = \{ \ \rightarrow$+over_lv(s3):[1,t],$\rightarrow$+over_lv(s3):[2,t],$\rightarrow$+over_lv(s3):[3,t],
$\rightarrow$+critical_lv(s3):[1,t],$\rightarrow$+critical_lv(s3):[2,t],$\rightarrow$+critical_lv(s3):[3,t],
$\rightarrow$+partial_block(s3):[1,t],$\rightarrow$+partial_block(s3):[2,t],$\rightarrow$+partial_block(s3):[3,t],
$\rightarrow$+total_block(s1,s2):[1,t],$\rightarrow$+total_block(s1,s2):[2,t],
$\rightarrow$+total_block(s1,s2):[3,t] $\}$

In order to compute the active semantics, we have to compute the fixpoint of $\Delta_{\Xi,sel_\mathcal{A}}(\langle \emptyset, EDB_\mathcal{A} \rangle)$. We let $I^e = EDB_\mathcal{A}$ and we start by computing $\Gamma_{\Xi,\emptyset}(I^e)$:

$\Gamma_{\Xi,\emptyset}(I^e) = I_1 = I^e \cup \{$ danger_lv(s1):[1,t],danger_lv(s2):[1,$\bot$],danger_lv(s3):[1,t],
danger_lv(s1):[2,$\bot$],danger_lv(s2):[2,t],danger_lv(s3):[2,t],
danger_lv(s1):[3,t],danger_lv(s2):[3,t],danger_lv(s3):[3,t],
danger_lv(s1):[$\{1,2\}$,t],danger_lv(s1):[$\{1,3\}$,t],danger_lv(s1):[$\{2,3\}$,t],
danger_lv(s2):[$\{1,2\}$,t],danger_lv(s2):[$\{1,3\}$,t],danger_lv(s2):[$\{2,3\}$,t],
danger_lv(s3):[$\{1,2\}$,t],danger_lv(s3):[$\{1,3\}$,t],danger_lv(s3):[$\{2,3\}$,t],
danger_lv(s1):[$\{1,2,3\}$,t],danger_lv(s2):[$\{1,2,3\}$,t],danger_lv(s3):[$\{1,2,3\}$,t],

---

[12] We recall that in Active U-Datalog, only the answers of the last simple transaction are returned (Bertino *et al.*, 1998).

```
                            tblock(s1,s2):[s,t],
                            ...   ...   ...
                            +over_lv(s3):[1,t],+over_lv(s3):[2,t],+over_lv(s3):[3,t],
                            +critical_lv(s3):[1,t],+critical_lv(s3):[2,t],+critical_lv(s3):[3,t],
                            +partial_block(s3):[1,t],+partial_block(s3):[2,t],+partial_block(s3):[3,t],
                            +total_block(s1,s2):[1,t],+total_block(s1,s2):[2,t],+total_block(s1,s2):[3,t]   }
```

Since $I_1$ is consistent, we let $\Delta_{\Xi,sel_\mathcal{A}}(\langle \emptyset, EDB_\mathcal{A}\rangle) = \langle \{\}, I_1\rangle$. The computation continues by computing $\Gamma_{\Xi,\emptyset}(I_1)$, obtaining the following set:

$\Gamma_{\Xi,\emptyset}(I_1) = I_2 = I_1 \cup \{$  pblock(s3):[s,t],
                            -partial_block(s3):[1,t],-partial_block(s3):[2,t],-partial_block(s3):[3,t]   $\}$

At this step, a conflict has been generated, since there is the request of both inserting and deleting atoms `partial_block(s3):[1,t], partial_block(s3):[2,t]`, and `partial_block(s3):[3,t]`. The conflicts related to these updates are the following:

- $C_1 = (1, partial\_block(s3), \{(r_1, \emptyset)\}, \{(r_2, \{X \leftarrow s1, Y \leftarrow s2, Z \leftarrow 1, W \leftarrow s3,$
  $V \leftarrow t\})\})$
  where $r_1$ is the rule $\rightarrow$ `+partial_block(s3):[1,t]` and $r_2$ is the rule `+total_block(X,Y):[Z,t], partial_block(W):[Z,V]`$\rightarrow$`-partial_block(W):[Z,V]`.
- $C_2 = (2, partial\_block(s3), \{(r_3, \emptyset)\}, \{(r_4, \{X \leftarrow s1, Y \leftarrow s2, Z \leftarrow 2, W \leftarrow s3,$
  $V \leftarrow t\})\})$
  where $r_3$ is the rule $\rightarrow$ `+partial_block(s3):[2,t]` and $r_4$ is the rule `+total_block(X,Y):[Z,t], partial_block(W):[Z,V]`$\rightarrow$`-partial_block(W):[Z,V]`.
- $C_3 = (3, partial\_block(s3), \{(r_5, \emptyset)\}, \{(r_6, \{X \leftarrow s1, Y \leftarrow s2, Z \leftarrow 3, W \leftarrow s3,$
  $V \leftarrow t\})\})$
  where $r_5$ is the rule $\rightarrow$ `+partial_block(s3):[3,t]` and $r_6$ is the rule `+total_block(X,Y):[Z,t], partial_block(W):[Z,V]`$\rightarrow$`-partial_block(W):[Z,V]`.

Since we assume that the conflict resolution functions privileges deletions, we set $B' = \mathsf{blocked}(I^e, \Xi, I_2, sel_\mathcal{A}) = \{(r_1, \emptyset), (r_3, \emptyset), (r_5, \emptyset)\}$ and we obtain $\Delta_{\Xi,sel_\mathcal{A}}(\langle\{\}, I_1\rangle) = \langle B', I^e\rangle$. We have now to compute $\Gamma_{\Xi,B'}(I^e)$, obtaining the following set:

$\Gamma_{\Xi,B'}(I^e) = I'_1 = I^e \cup \{$  danger_lv(s1):[1,t],danger_lv(s2):[1,$\bot$],danger_lv(s3):[1,t],
                            danger_lv(s1):[2,$\bot$],danger_lv(s2):[2,t],danger_lv(s3):[2,t],
                            danger_lv(s1):[3,t],danger_lv(s2):[3,t],danger_lv(s3):[3,t],
                            danger_lv(s1):[{1,2},t],danger_lv(s1):[{1,3},t],danger_lv(s1):[{2,3},t],
                            danger_lv(s2):[{1,2},t],danger_lv(s2):[{1,3},t],danger_lv(s2):[{2,3},t],
                            danger_lv(s3):[{1,2},t],danger_lv(s3):[{1,3},t],danger_lv(s3):[{2,3},t],
                            danger_lv(s1):[{1,2,3},t],danger_lv(s2):[{1,2,3},t],danger_lv(s3):[{1,2,3},t],
                            tblock(s1,s2):[s,t],
                            ...   ...   ...
                            +over_lv(s3):[1,t],+over_lv(s3):[2,t],+over_lv(s3):[3,t],
                            +critical_lv(s3):[1,t],+critical_lv(s3):[2,t],+critical_lv(s3):[3,t],
                            +total_block(s1,s2):[1,t],+total_block(s1,s2):[2,t],+total_block(s1,s2):[3,t]
                            $\}$

Since $I'_1$ is consistent, we let $\Delta_{\Xi,sel_\mathcal{A}}(\langle B', I^e\rangle) = \langle B', I'_1\rangle$. The computation continues by computing $\Gamma_{\Xi,B'}(I'_1)$, obtaining the following set:

$\Gamma_{\Xi,B'}(I'_1) = I'_2 = I'_1 \cup \{$  pblock(s3):[s,t],
                            -partial_block(s3):[1,t],-partial_block(s3):[2,t],-partial_block(s3):[3,t]
                            $\}$

Since $I'_2$ is consistent, we let $\Delta_{\Xi, sel_\mathcal{A}}(\langle B', I'_1 \rangle) = \langle B', I'_2 \rangle$. It is possible to prove that additional iterations do not generate any new constrained atom, therefore $\Delta^\omega_{\Xi, sel_\mathcal{A}}(\langle \{\}, EDB_\mathcal{A} \rangle) = \langle B', I'_2 \rangle$. From this, we obtain

$$Sem_{P, sel_\mathcal{A}}(T) = \langle X \leftarrow s3, Y \leftarrow s1, Z \leftarrow s3, EDB'_\mathcal{A}, \mathsf{Commit} \rangle$$

where $EDB'_\mathcal{A} = \mathsf{incorp}(I'_2, EDB_\mathcal{A})$ and:

$\mathsf{incorp}(I'_2, EDB_\mathcal{A}) = \{$ sensor(s1):[1,t],sensor(s2):[1,$\perp$],sensor(s3):[1,t],over_lv(s1)[1,t],
critical_lv(s1):[1,t],sensor(s1):[2,$\perp$],sensor(s2):[2,t],sensor(s3):[2,t],
over_lv(s2)[2,t],critical_lv(s2):[2,t],sensor(s1):[3,t],sensor(s2):[3,t],
sensor(s3):[3,t],over_lv(s3):[1,t],over_lv(s3):[2,t],over_lv(s3):[3,t],
critical_lv(s3):[1,t],critical_lv(s3):[2,t],critical_lv(s3):[3,t],
total_block(s1,s2):[1,t],total_block(s1,s2):[2,t],total_block(s1,s2):[3,t]
$\}$

$\diamond$

## 6 Conclusions and future work

In this paper, we defined a logical framework for modeling queries, updates, and update propagation against a set of heterogeneous knowledge bases. The framework has been obtained by extending the amalgamated knowledge base framework proposed in (Subrahmanian, 1994) to deal with updates and updates propagation. To this purpose, the local databases and the mediator have been modeled as Annotated Active U-Datalog databases. In this way, each local source and the mediator are composed of a set of ground facts, a set of deductive rules and a set of active rules, whose semantics has been defined according to the PARK semantics proposed in (Gottlob *et al.*, 1996). A fixpoint semantics for the proposed language has also been proposed, extending those presented in (Subrahmanian, 1994) and in (Bertino *et al.*, 1998).

This work can be extended in several ways. A first important question concerns the definition and analysis of properties concerning the execution of distributed queries, transactions and active rules. Another important issue is the extension of AmAU-Datalog with negation in deductive rules and the definition of proper semantics. Finally, an additional important direction concerns the extension of the proposed language to model not only integration but also cooperation among the various sources. To this purpose, we plan to integrate the capabilities of the proposed framework with those of Heterogeneous U-Datalog (Bertino *et al.*, 2000), by providing each local sources with the ability to communicate and exchange information with other sources. The result would be a framework for integrating knowledge bases in a fully static, dynamic, and cooperative way.